\documentclass[aps,pra,twocolumn,superscriptaddress,longbibliography]{revtex4-1} 
\usepackage{subfigure}
\usepackage{hyperref}
\usepackage[pdftex]{color}
\usepackage{soul}

\usepackage{graphicx}
\usepackage{caption}
\usepackage{amsmath}
\usepackage{amssymb}
\usepackage{mathtools}
\usepackage{dsfont}
\usepackage{bm}
\usepackage{color}
\usepackage{appendix}
\usepackage[normalem]{ulem}

\usepackage{enumitem}
\usepackage{amsthm}
\usepackage{siunitx}
\usepackage[usenames,dvipsnames]{xcolor}
\usepackage{multirow}

\numberwithin{theorem}{subsection}

\newcommand{\bqa}{\begin{eqnarray}}
\newcommand{\eqa}{\end{eqnarray}}
\newcommand{\beq}{\begin{equation}}
\newcommand{\eeq}{\end{equation}}

\hypersetup{
    colorlinks=true,       
    linkcolor=cyan,          
    citecolor=magenta,        
    filecolor=magenta,      
    urlcolor=cyan,           
    runcolor=cyan
}

\begin{document}

\title{Advantage of pausing: parameter setting for quantum annealers}
\author{Zoe Gonzalez Izquierdo}
\affiliation{QuAIL, NASA Ames Research Center, Moffett Field, California 94035, USA}
\affiliation{USRA Research Institute for Advanced Computer Science, Mountain View, California 94043, USA}
\author{Shon Grabbe}
\affiliation{QuAIL, NASA Ames Research Center, Moffett Field, California 94035, USA}
\author{Husni Idris}
\affiliation{Aviation Systems Division NASA Ames Research Center, Moffett Field, California 94035, USA}
\author{Zhihui Wang}
\affiliation{QuAIL, NASA Ames Research Center, Moffett Field, California 94035, USA}
\affiliation{USRA Research Institute for Advanced Computer Science, Mountain View, California 94043, USA}
\author{Jeffrey Marshall}
\affiliation{QuAIL, NASA Ames Research Center, Moffett Field, California 94035, USA}
\affiliation{USRA Research Institute for Advanced Computer Science, Mountain View, California 94043, USA}

\author{Eleanor Rieffel}
\affiliation{QuAIL, NASA Ames Research Center, Moffett Field, California 94035, USA}
\date{\today} 

\begin{abstract}
Prior work showed the efficacy of pausing midanneal: such a pause improved the probability of success by orders of magnitude in a class of native problem instances and improved the time to solution in a class of embedded problem instances. A physics-based picture provides qualitative suggestions for where pausing midanneal is effective, for the interplay between annealing schedule parameters and other annealing properties and parameters such as embedding size and strength of the ferromagnetic coupling $|J_F|$, and for the conditions under which pausing can improve the time to solution. Here, through demonstrations on an updated annealing architecture that has higher connectivity than previous annealers, and on multiple embedded problem classes, we are able to confirm various aspects of this picture.
We demonstrate the robustness of the optimal pause parameters across platforms and problem classes, explore how to set $|J_F|$ to optimize performance in different scenarios, and provide empirical evidence that short pauses trump longer overall annealing times in time to solution. We also identify the number of different coefficients in a problem as a predictor of problem hardness, and explore its interplay with the optimal $|J_F|$ and embedding size.
Based on these results we are able to present qualitative guidelines for parameter setting in quantum annealers.

\end{abstract}
\maketitle

\section{INTRODUCTION}
\label{sec:intro}

The theory behind quantum annealing has now existed for over two decades~\cite{Finnila,Farhi-98,Kadowaki,Farhi,Aharonov-04,Santoro}, and actual physical quantum annealers for more than one~\cite{Harris2010,Harris2010_2,Johnson2011quantum,Boixo2013experimental,Bunyk2014,Jiang2018,Hauke2020}. Yet, the much sought-after speedup for solving optimization problems with these devices has remained elusive~\cite{Ronnow2014defining,Boixo2014evidence,katzgraber2015seeking,Mandra2018,Pearson2019}. 
One way of improving performance for optimization problems is pausing midanneal~\cite{Marshall19_Pausing,Venturelli2019,p-spin-pause,Vinci2020,chen_pausing,gonzalez2021}. By making clever use of the thermalization that takes place in finite-temperature annealers and taking into consideration the dynamics of the system throughout the annealing process, this technique was first shown to increase the ground-state (GS) success probability, $p_{success}$, for native problems~\cite{Marshall19_Pausing}, and later confirmed to work for embedded problems and adjusted to also improve the time to solution~\cite{gonzalez2021}. This makes pausing a necessary ingredient of benchmarking studies going forward.

A quantum annealing solver has multiple parameters or variables whose values can be set by the user, including choice of annealing schedule, total anneal time and embedding parameters. Adding a pause further complicates this landscape as two more parameters (location and duration of the pause) come into play. Guidance in what regime to look for the optimal or good parameters is therefore valuable for efficiently boosting success probability. Benchmarking studies that help inform parameter setting for families of problems have been a very useful tool to navigate these questions~\cite{Rieffel14CaseStudy,Vinci2016,Albash2018,Quiroz2019,Grant2021}, and they become crucial as quantum annealers offer further opportunities to be controlled by user input. 

As the landscape becomes more complicated, exploring a large range of values and combinations becomes prohibitively costly. The aspiration to make annealers useful for a multitude of applications adds to the issue, with understanding how certain characteristics of the problems interact with annealing parameters coming into play. Moreover, as newer and more diverse quantum annealing devices become available~\cite{Weber2017}, we need to consider how their potential differences affect performance and parameter setting.

We explore these questions in the present work, by testing several predictions on optimal parameter regions across two annealing devices and three optimization problems, to help provide parameter setting guidance in a variety of scenarios.
In particular, in this paper we do the following:

\begin{itemize}
    \item Show through demonstrations on quantum annealing hardware that a short pause midanneal, in the region of pause location~\footnote{For definitions of the pause location and other concepts related to quantum annealing, see Sec.~\ref{sec:background}.} $s_p=0.3$-0.5,
    can improve time to solution for ensembles of three different non-native problems as well as for two quantum annealing devices with different architectures, further confirming that the general location at which a pause will help remains consistent with the physical picture describing the dynamics at different times in the anneal~\cite{gonzalez2021}.
    \item Confirm that within that region, larger embedding sizes and larger ferromagnetic couplings will lead to an earlier optimal location, and vice versa.
    \item For a problem class with optimal annealing time higher than the hardware minimum, demonstrate that introducing a short pause brings the optimal down to at most the hardware minimum.
    This provides evidence for the fact that extra time is only helpful in some regions of the anneal, and adds support to the idea that, even with access to shorter annealing times, pausing will still be beneficial.
    \item Provide evidence for coefficient heterogeneity (i.e. the number of different $h_i$ and $J_{ij}$ values in the logical problem) being a strong predictor of problem hardness (partly due to hardware precision limitations). We do this by considering three problems whose QUBOs have different degrees of coefficient heterogeneity, which gets amplified through the potential coefficient splits over multiple qubits and couplers that occur during embedding.
    \item Show the more secondary role that problem size plays in problem hardness, finding its effect to become apparent once coefficient heterogeneity stays relatively constant.
    \item Qualify the correlation between vertex model\footnote{We use vertex model to refer to the set of physical qubits representing a single logical variable after minor-embedding. The word ``chain'' is sometimes used in the literature with the same meaning, but we prefer vertex model given that their structure is often not chain-like.} size and optimal $|J_F|$, showing that it gets complicated by coefficient heterogeneity, rather than simply larger vertex models benefiting from stronger couplings. 
    \item Through the above points, provide insights for parameter setting across diverse classes of problems and hardware architectures, both general and more specific based on knowledge of the logical and embedded problem, thus removing the need to explore all regions when trying to optimize performance for optimization problems.
    \item Define an information sharing problem within the context of collective autonomous mobility~\cite{Idris20} as one of the three problems analyzed in this paper, and formulate it as a QUBO. In this problem, information sharing among vehicles is optimized to mitigate communication bandwidth constraints in high density traffic operations such as urban air mobility~\cite{Thipphavong18}.
\end{itemize}

The rest of the paper is organized as follows. In Sec.~\ref{sec:background} we give a brief overview of quantum annealing. Further technical details including a description of the two devices we use can be found in App.~\ref{appendix:device_details}. Sec.~\ref{sec:methods} introduces the three optimization problems (for which QUBO mappings can be found in App.~\ref{appendix:mapping_mst},~\ref{appendix:mapping_gc}, and~\ref{appendix:mapping_info}), as well as the metrics and parameters used in the demonstration. We present results for the same set of instances solved on two different devices in Sec.~\ref{subsec:device_comp_results}, while Sec.~\ref{subsec:problem_comp} contains results for the three different problems solved on the newer device. The findings from these results, as well as future research avenues are discussed in Sec.~\ref{sec:discussion}. A summary of our main points and closing remarks can be found in Sec.~\ref{sec:conslusions}.

\section{BACKGROUND: QUANTUM ANNEALING}
\label{sec:background}

Quantum annealing is a quantum metaheuristic for optimization, and quantum annealers are quantum hardware designed to run this metaheuristic. Any classical cost function $C(x)$ that is a polynomial over binary variables $x\in\{0,1\}^n$ can, with the addition of auxiliary variables, be turned into a quadratic cost function. Problems with quadratic cost functions over binary variables without additional constraints are called quadratic unconstrained binary optimization (QUBO) problems, and are solvable by quantum annealers (subject to size constraints imposed by each particular device).

Quantum annealing relies on the fact that the solution to such an optimization problem can also be understood as the GS of an Ising problem Hamiltonian~\cite{Morita2008} $H_p = \sum_{<ij>} J_{ij} \sigma_i^z \sigma_j^z + \sum_i h_i \sigma_i^x$, where $\sigma_i^x$ and $\sigma_i^z$ are individual Pauli matrices acting on spin $i$, $J_{ij}$ represents the strength of the coupling between spins $i$ and $j$, and $h_i$ that of the bias on spin $i$. Combinatorial optimization problems can be expressed in Ising form by a straightforward mapping between QUBO and Ising (mapping binary variables 0 and 1 to spin variables $\pm 1$).

Then, to find the GS of $H_p$ (and thus the solution to the corresponding QUBO) quantum annealing is carried out by evolving the system under a time-dependent Hamiltonian
$H(s) = A(s)H_d + B(s)H_p$, 
where $H_d$ is a driver Hamiltonian (most commonly a transverse field $H_X = -\sum_i \sigma_i^x$), $s$ is a dimensionless time parameter that ranges from $0$ to $1$ over the course of a single anneal, and $A(s)$ and $B(s)$ are device-dependent functions determining the strength of the driver and problem Hamiltonians, respectively. These functions are such that $A(0) \gg B(0) \approx 0$ and $B(1) \gg A(1) \approx 0$, so that the time-dependent Hamiltonian evolves from $H_d$ to $H_p$ throughout the anneal. The system is then initialized in the easy to prepare GS of $H_d$, and expected to remain near the GS of $H(s)$ throughout, to finally yield the GS of $H_p$ (or equivalently, the solution to the optimization problem) upon final measurement. More information about quantum annealing generally, including mappings of optimization problems to QUBO can be found in \cite{Choi19, Rieffel14CaseStudy,STMapping}.

We use two different quantum annealing devices for our demonstrations: D-Wave 2000Q (DW2K), and D-Wave Advantage (DWA), with the latter being the newer of the two. Some technical details of these devices can be found in Appendix~\ref{appendix:device_details}.

\subsection{Parameter setting: schedules and ferromagnetic coupling}
\label{subsec:param_setting}

While the functions $A(s)$ and $B(s)$ cannot be changed by the user in currently available devices, the dependence $s(t)$ can be modified under certain constraints: $s$ must be a linear, nondecreasing function of $t$, starting at 0 and ending at 1, with a maximum slope of $1 \mu s^{-1}$, and a limited number of slope changes allowed. This leads to different so-called schedules $s(t)$. In particular, we use schedules that include a pause, that is, a period during which $ds/dt = 0$, meaning that $H(s)$ is held constant and the anneal is effectively paused. 

For most application problems, since the hardware has restricted qubit connectivity, the resulting QUBO is unlikely to conform to this hardware connectivity, and a one-to-one correspondence between logical variables and physical qubits does not exist. Instead, we must perform minor embedding~\cite{Choi}, which enables coupling between logical variables in the QUBO graph by representing each of them by an appropriate set of physical qubits that will allow the required connectivity. Following standard terminology in graph theory, each such set of physical qubits is called a {\it vertex model} for its corresponding logical variable.

Because all qubits in the vertex model should act as a single variable (i.e. be aligned, otherwise the vertex model is considered to be broken, and the solution containing it will be discarded), they are ferromagnetically coupled to promote this collective behavior. We use the same coupling strength $|J_F|$ for all the couplings within a vertex model ($J_F$ is always negative, so we typically refer to its magnitude $|J_F|$). Problems that do not require embedding because their structure matches that of the hardware are called {\it native problems} for that hardware.

While $|J_F|$ can be set to a large value such that the embedded problem preserves the GS of the logical problem, 
and analytical bounds on this value can be obtained~\cite{Choi}, too large a $|J_F|$ can reduce quantum annealing performance. 
Physically there is an energy limit on the Hamiltonian as a whole, and too large a $|J_F|$ relative to other parameters would mean that all of the problem parameters could reduce performance due to precision issues and noise in implementation. 
Furthermore, the energy spectrum throughout the anneal varies with the value of $|J_F|$, and its effect on the annealing often requires careful case-by-case consideration~\cite{Choi19, Rieffel14CaseStudy, Warburton, Venturelli15, marshall2020perils, perils2}.
Thus, optimally setting the ferromagnetic coupling $|J_F|$ is a challenging task. Prior work has shown there is a sweet spot for this value.
Physically this makes sense because a stronger $|J_F|$ makes it less likely for individual qubits within a vertex model to flip, which helps to avoid breaking the vertex model, but too large a $|J_F|$ makes it increasingly costly for the vertex model qubit values to flip together, potentially preventing the system from leaving a nonoptimal configuration. To boost the probability of success, $|J_F|$ must strike the right balance, leading to better chances of arriving at---and staying in---the correct configuration.

The schedule $s(t)$ can also significantly affect performance. Of particular interest to us are schedules that include a pause where for some subinterval $s(t)$ is constant (i.e., $H(s)$ is constant for a specified time). 
It was first observed for an ensemble of native problems that a pause at a location (generally) insensitive to the instance specifics boosts the success probability by orders of magnitude~\cite{Marshall19_Pausing}, and a physical picture was presented explaining this effect. It was later shown that it also applies to embedded problems, and that with the right pause location and duration~\cite{Albash2021} the time to solution can also be improved~\cite{gonzalez2021}.

\section{METHODS}
\label{sec:methods}

\subsection{Problem statements}
\label{subsec:problems}

\subsubsection{Minimum spanning tree with bounded degree}
\label{subsubsec:MST}

For our comparison across devices, we solve the minimum spanning tree with bounded degree (BD MST) problem. The results for the older of the two devices correspond to the results from Ref.~\cite{gonzalez2021}. Spanning trees are useful for a number of reasons, including the designing of efficient routing algorithms and for their wide-ranging applications to network design, cluster analysis \cite{Grygorash06} and bioinformatics \cite{Xu02}. The problem statement is as follows: given a connected, undirected graph $G=(V,E)$ with edge weights $w_{uv}$, $(uv)\in E$, and an integer $\Delta\ge 2$,  find a minimum weight spanning tree of maximum degree at most~$\Delta$. A spanning tree of $G$ is a subgraph of $G$ that is a tree and contains all vertices of~$G$. The tree weight to be minimized is $\sum_{(uv) \in T} w_{uv}$. 

Without the bound on the degree, determining if there exists a spanning tree of weight $W$ for a graph $G$ can be decided in polynomial time, and different efficient algorithms exist to find a minimum weight tree~\cite{Cormen-01}. However, once we add the constraint on the degree, the problem becomes NP complete~\cite{GareyJohnson}. We use a level-based QUBO mapping that can be found in Appendix~\ref{appendix:mapping_mst}.

\subsubsection{Graph coloring}
\label{subsubsec:gc}

Our second problem is graph coloring (GC). Given an undirected graph $G = (V, E)$ with $n$ nodes, and $k$ colors, the objective is to find a color assignment for the nodes such that no nodes with the same color are connected by an edge. Our instances are randomly generated 4-regular graphs with $n=12$, 14 and 16, and $k=5$ colors, and we use a standard one-hot encoding for the mapping to QUBO, described in Appendix~\ref{appendix:mapping_gc}.

\subsubsection{Information sharing}
\label{subsubsec:info}

The third problem we consider is an information sharing (INFO) problem within the context of collective autonomous mobility~\cite{Idris20} that will assist with high-density urban air mobility~\cite{Thipphavong18}. 

\begin{figure}[h]
\includegraphics[width=\linewidth]{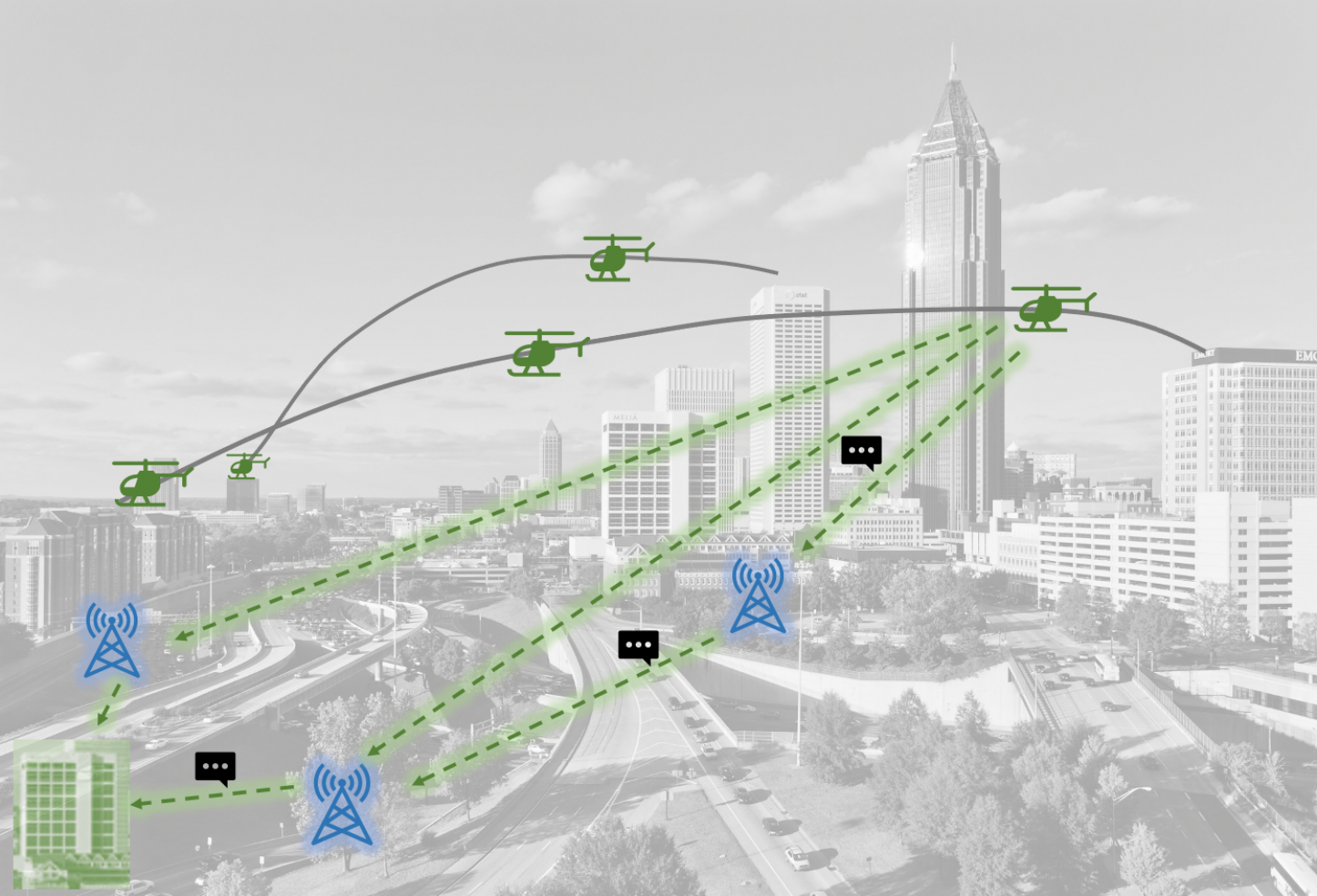}
  \caption{{\bf Information sharing illustrative example.} 
  Three messages (black symbols) are transmitted from an airborne vehicle (upper right) to a single ground-based facility (lower left) via two network communication towers (blue cell-tower symbols). The vehicles, ground-based facilities, and cell towers correspond to the nodes in the graph associated with the problem, while the green dashed lines between them are the edges, and represent the possible paths for message transmission. Though not relevant to our problem, the solid gray lines depict the trajectories flown by the vehicles.}
  \label{fig:info_sharing}
\end{figure}

A set of messages are to be transmitted over a communications network.  The network will have one or more senders (e.g. an airborne vehicle), one or more receivers (e.g a ground-based facility), as well as intermediate points (e.g. cell towers). These elements can be represented as the nodes of a graph, with the paths between them acting as edges. Each edge will have an associated weight, equal to the (finite, integer) time it takes a message to traverse it. An illustrative example of this scenario is depicted in Fig.~\ref{fig:info_sharing}.

Each message has an assigned path of transmission through the network---with a sender, a receiver, and some intermediate points---as well as a scheduled emission time. The capacity of the network is finite (and might not be the same at every point). This, combined with the fact that transmission takes finite time, means that sending all the messages at their scheduled emission times might not be possible (as capacity could be exceeded at certain locations and times). To respect this limitation, the transmission of certain messages can be delayed with respect to the scheduled time, with the caveat that this delay carries some cost per unit time.

Each message has associated a cost of transmission delay per unit time. It can be understood as a representation of the message's priority level, and the relevance of it being delivered on time. To solve the problem, each message transmission must be assigned a time delay in such a manner that the total cost of delay is minimized. If every message can be transmitted at its scheduled emission time, the total cost is zero. If delays are required, there will be some finite cost.

Certain messages may be time critical, due to safety or other reasons (which can be represented by their cost of delay being much greater than the rest). The paths for such messages will be hard coded into the problem setup, by reducing the network capacity along their path, so that no delay can be assigned to them and other messages cannot use the network bandwidth that has been preassigned to the high priority ones.

The full set of problem parameters, along with the mapping to QUBO, is provided in Appendix~\ref{appendix:mapping_info}.

So far, we consider the transmission paths of the messages to be predetermined. However, depending on the particular communications network, several paths could exist between the sender and receiver of any given message, and the total cost of delay could potentially be lowered by allowing changing of certain paths.
While we focus on solving the delay optimization aspect with the quantum annealer in this study, both delays and paths could be optimized by implementing a hybrid algorithm, alternating classical and quantum routines (for an example of a similar problem, in the context of air traffic, where paths are optimized, see Ref.~\cite{Sadovsky2014}). The full algorithm would have the following steps:

\begin{enumerate}
    \item Taking as inputs the set of messages, the graph representing the communications network, and the source and destination nodes for each message, the shortest path for each transmission is calculated using a standard classical algorithm.
    \item Using the calculated shortest paths, the quantum annealer finds the set of transmission delays that result in the minimum total cost of delay.
    \item The costliest delay is identified and the path of that message changed, avoiding the location and time where the network capacity was exceeded, prompting a delay.
    \item With this new path, the cost optimization problem is solved again in the quantum annealer.
    \item If the total cost is reduced, the new path is accepted, and the path-changing routine repeated with the new costliest delay. If the total cost increases, it can be for one of two reasons. If it is due to the new path resulting in a later arrival time, and thus a more costly delay than the one incurred by the original path, the path is not changed; instead, the second costliest delay is considered and the path-changing routine repeated there. If, on the other hand, the worse outcome is due to this new path causing a previously nonexistent delay for a different message, the change is kept and the path-changing routine applied to the path of the message affected by the new delay.
    \item The path-changing routine is repeated until all delays have been considered for a path change.
\end{enumerate}

\subsection{Parameters and metrics}
\label{subsec:metrics}

We use the empirical probability of success ($p_{success}$) and time to solution ($T_S$) as our figures of merit for determining how likely a problem is to be solved, defined as:
\begin{align}
    & p_{success} = \frac{\text{no. anneals with correct solution}}{\text{total no. anneals}},
\\
& T_S = \frac{\log(1 - 0.99)}{\log(1.0 -  p_{success})}t_{tot}\;,
\label{eq:TTS}
\end{align}
where the total time $t_{tot}=t_a+t_p$ is the time spent on each anneal, taking into account both the annealing time $t_a$ and, in the case of schedules with a pause, the pause duration $t_p$. 

These two measures are complementary to each other. The $T_S$ figure of merit reports the expected time required to solve the problem with $99\%$ confidence.
While $p_{success}$ is directly determined by and hence provides a portal to understand the underlying physical process, $T_S$ gives a more practical measure that is universal across different parameter ranges and different solvers.
A higher success probability does not necessarily mean a lower $T_S$.  For instance, we might get a slightly higher $p_{success}$ by using a longer annealing time $t_a = 100 \mu$s than with a shorter one $t_a = 1 \mu$s, 
yet the chance of finding the solution might be higher by repeating the $t_a = 1 \mu$s runs 100 times than by doing the 
$t_a = 100 \mu$s anneal once.
It is for this reason that when optimizing the different parameters for solving a problem, we aim to minimize $T_S$ rather than maximize $p_{success}$.

We consider only that the correct solution has been found when the optimal is returned. For GC instances, this always means that the total energy returned by D-Wave is 0, as any nonzero contributions would come from either a node not being assigned exactly one color or from adjacent nodes being assigned the same color. For the BD MST and INFO problems, each instance will have a particular minimum energy which might be nonzero, and comes entirely from the cost-function contribution, without violating any penalty terms. All these instances are small enough to be solved either by inspection or with a classical routine, and we verify whether the minimum energy returned by D-Wave is in fact that of the correct solution. We also ensure that all the contributions from the penalty terms are 0.

We discard any potential solutions returned by D-Wave with inconsistent or broken vertex models, i.e., where not all the qubits within a given vertex model are aligned.

Unless otherwise noted, all our results are obtained in the following manner: for each set of annealing parameters and each problem instance, we perform 100 gauges (or partial gauges if $|J_F|>1$, where the gauge is only applied to couplings $\leq 1$), with 500 anneals each, for a total of 50,000 anneals (or reads) per run. The number of correct solutions found in those 50,000 anneals is then divided by the total number of anneals to obtain $p_{success}$ for each instance. Then, a bootstrap procedure is performed over the ensemble of instances, by drawing a number of samples equal to the number of instances (with replacement), from which a median is obtained, and repeating this process $10^5$ times to finally calculate a median of medians, which is reported as our data point. The 35th and 65th percentile values are used for the error bars.

For the standard no pause schedules, we can vary the ferromagnetic coupling $|J_F|$ (sometimes referred to as chain strength in the literature) as well as the annealing time $t_a$. For schedules with a pause, we consider $|J_F|$ and $t_a$, and also the pause location within the anneal, $s_p$, and its duration $t_p$.

After mapping to QUBO, all our instances need to be minor embedded to fit the adjacency graph of the device. We use the standard embedding heuristic implemented by D-Wave software, which we run ten times and keep the smallest embedding found (i.e. with the smallest number of physical qubits). Embedding size is one of several factors that affect performance, as we discuss in Sec.~\ref{sec:discussion}, and selecting a smaller embedding does not necessarily guarantee a higher probability of success. However, it is a straightforward metric which does not require additional running time, and we find it can improve performance for larger and harder problems.

\section{RESULTS}
\label{sec:results}

We present our results in this section, split into the study of a single problem on two quantum annealers (Sec.~\ref{subsec:device_comp_results}) and that of three different problems on a single annealer (Sec.~\ref{subsec:problem_comp}). We limit ourselves here to laying out our results along with some brief comments, while a more in-depth discussion of their meaning and implications is reserved for Sec.~\ref{sec:discussion}.

\subsection{Comparison across devices}
\label{subsec:device_comp_results}

We study an ensemble of 45 instances of BD MST problems on five nodes that are chosen by exhausting all connected graphs with $n=|V|=5$. Weight sets are uniformly drawn from 1 to 7. Graphs and weight sets are combined to yield these 45 unique instances. Once mapped to QUBO following Appendix~\ref{appendix:mapping_mst}, they have between 32 and 74 logical qubits. All instances require embedding on both devices. Table~\ref{tab:mst} shows the embedded size differences between DW2K and DWA, with the latter providing an approximately 2.5-fold improvement in size, as expected from its 2.5-fold increase in connectivity (degree 6 versus degree 15) compared with the former. Results for DW2K were obtained from our previous study~\cite{gonzalez2021}.

\begin{table}[ht]
\begin{tabular}{| c | c | c | c | c |}
\hline
  & Physical size & \begin{tabular}{@{}c@{}}Median \\ vertex model \end{tabular}  & Optimal $|J_F|$ \\
 \hline
 DW2K & 83-485 & 1.5-7 & 1.6 \\
 \hline
 DWA & 38-188 & 1-2 & 0.8 \\ 
 \hline
\end{tabular}
\caption{BD MST embedding data}
\label{tab:mst}
\end{table}

\begin{figure}[h]
\includegraphics[width=\linewidth]{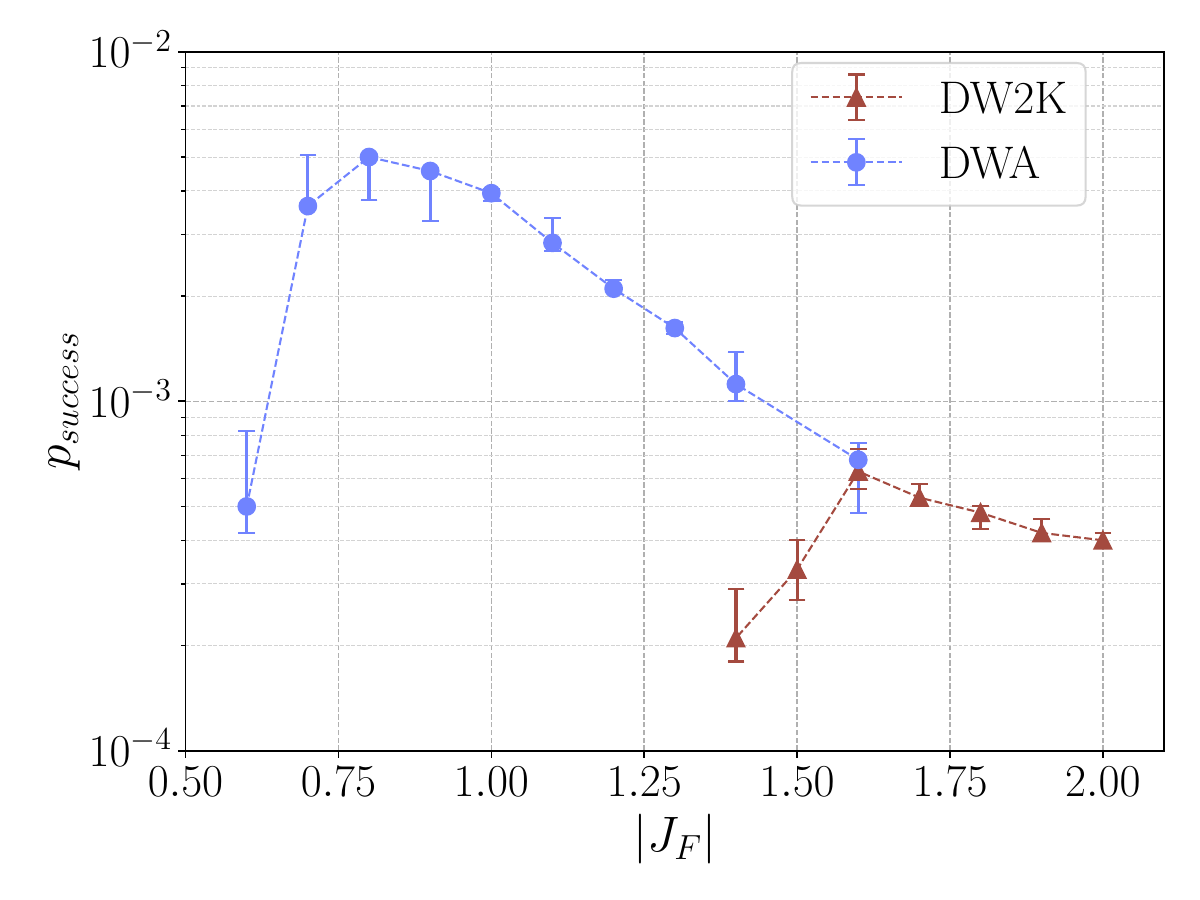}
  \caption{{\bf Difference in $p_{success}$ and optimal $|J_F|$ between the two devices.} 
  $p_{success}$ for an ensemble of 45 BD MST instances with $n=5$, annealing without a pause. The best performance for DW2K is seen at $|J_F|$=1.6, while DWA has its optimal at $|J_F|=0.8$. The maximum $p_{success}$ attained is also much higher for the newer DWA.}
  \label{fig:adv_vs_2kq_no_pause}
\end{figure}

We start by finding the optimal ferromagnetic coupling when using a standard annealing schedule (without a pause). We choose an annealing time of 1 $\mu$s. This is shown in Fig.~\ref{fig:adv_vs_2kq_no_pause}. The optimal ferromagnetic couplings differ significantly between devices. DW2K does best at $|J_F|=1.6$, while for DWA it is $|J_F|=0.8$. Given that DW2K has much larger vertex models, keeping them from breaking requires a stronger coupling (we discard any solutions returned with broken vertex models). With the smaller vertex models on DWA, the exploration of configuration space provided by a weaker coupling outweighs the higher probability of breaking vertex models, as evidenced by a $7.5\times$ improvement in $p_{success}$ from DW2K to DWA, when considering them at their respective optimal $|J_F|$. 

Optimization of $|J_F|$ is crucial for getting the best possible outcome. It is highly dependent on embedding, which makes it change not only between devices and problems, but also between instances and even between different embeddings of the same instance, as we discuss in Sec.~\ref{subsec:problem_comp}. For example, if we use the DW2K optimal value with DWA, their $p_{success}$ would appear to be comparable. Also, because the instances in the ensemble each have their own optimal $|J_F|$, while $|J_F|=0.8$ is best overall on DWA, three of the 45 instances do not solve for this value; one of them solves for $|J_F|=0.9$ and the other two for $|J_F|=1.1$. We discuss this behavior in Sec.~\ref{sec:discussion}. On the other hand, DW2K was unable to solve seven of the 45 instances regardless of $|J_F|$.

\begin{figure}[h]
\includegraphics[width=\linewidth]{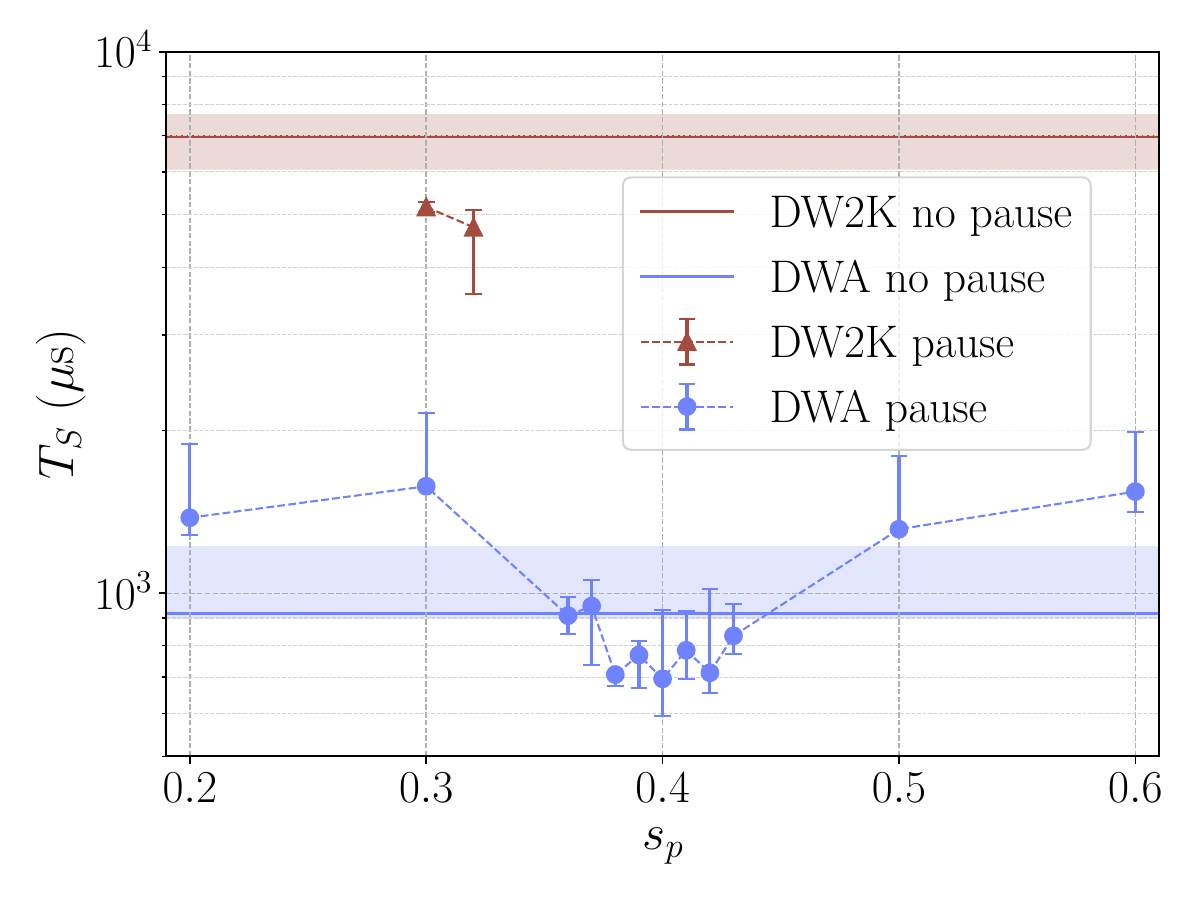}
  \caption{{\bf Improvement of $T_S$ with the introduction of a pause for the two devices.} 
  $T_S$ for an ensemble of 45 BD MST instances with $n=5$, annealing with a pause at the location indicated on the $x$ axis, and pause duration 1 $\mu$s for DW2K and 0.2 $\mu$s for DWA. In both cases $T_S$ is improved by about $20\%$ when optimizing location and duration of the pause.}
  \label{fig:adv_vs_2kq_pause}
\end{figure}

We also compare the case of a schedule with a pause, which, in our previous study, we found by carefully choosing its location and duration was able to improve not only $p_{success}$ but also $T_S$. The beneficial pausing region will be affected by the difference in the $A(s)$ and $B(s)$ functions (which would shift the region earlier) as well as by the smaller vertex models and weaker $|J_F|$ (both of which would shift it later). As shown in Fig.~\ref{fig:adv_vs_2kq_pause}, we find that the latter effect outweighs the former, with the optimal pause location going from $s=0.3$-$0.32$ in DW2K to $s=0.38$-$0.4$ in DWA. The improvement with $T_S$ for the ensemble is about $20\%$ for both devices. Only two data points are shown for DW2K because data for other locations was not obtained for the full ensemble. The pause duration for DW2K is 1 $\mu$s, which is found to be optimal in the range $[0.25, 2] \mu$s (although $t_p=0.5$ and 2 $\mu$s are within margin of error), and for DWA it is 0.2 $\mu$s, with no statistically significant differences in the range $[0.15, 2] \mu$s, but worse performance observed at 5 $\mu$s (no longer improving upon no pause results).

Finally, we include in Appendix~\ref{appendix:mst_n6} additional results for an ensemble of instances on six nodes (instead of five). None of them are solvable on DW2K, so these are only for DWA.

\subsection{Comparison across problems}
\label{subsec:problem_comp}

We now explore how different problems perform on the same device, for which we choose the newer generation DWA. In this section we limit ourselves to presenting a summary of our results, while a more thorough discussion can be found in Sec.~\ref{sec:discussion}. 

The ensemble of BD MST instances is the same as described earlier in Sec.~\ref{subsec:device_comp_results}. The GC instances are random 4-regular graphs on 12, 14, and 16 nodes, with five colors. For each of the three sizes we generate 20 different instances, and ten more are used for $n=16$ in some of the runs when explicitly stated, for a total of 30. The INFO ensemble is comprised of nine instances, with between two and four messages being transmitted. The graph representing the possible transmission paths has six nodes arranged in a $3\times 2$ lattice, with edge weights (i.e. transmission times between nodes) being one or two time units, and capacities of one or two messages. Paths consist of three or four edges, cost of delays is one to three units, and number of top priority messages 0 or 1. Detailed information about logical and physical sizes of these instances is displayed in Table~\ref{tab:problem_size}.

\begin{table*}[ht]
\begin{tabular}{| c | c | c | c | c | c | c | c |}
\hline
  \multicolumn{2}{| c |}{Problem class} & No. instances & Logical size & Physical size & \begin{tabular}{@{}c@{}}Mean \\ physical size \end{tabular}  &  \begin{tabular}{@{}c@{}}Mean \\ vertex model \end{tabular} & Mean degree \\
 \hline
 \multicolumn{2}{| c |}{BD MST} & 45 & 32-74 & 38-188 & 81 $\pm$ 22 & $1.7 \pm 0.3$ & $4.6 \pm 0.3$ \\
 \hline
 \multirow{3}{*}{GC} & $n=12$ & 20 & 60 & 133-180 & $166 \pm 11$ & $2.8 \pm 0.2$ & $5.1 \pm 0.2$ \\
 \cline{2-8}
    & $n=14$ & 20 & 70 & 144-219 & $193 \pm 16$ & $2.8 \pm 0.2$ & $5.2 \pm 0.3$ \\ 
\cline{2-8}
    & $n=16$ & 20-30 & 80 & 162-269 & $236 \pm 20$ & $3.0 \pm 0.2$ & $5.0 \pm 0.3$ \\ 
 \hline
 \multicolumn{2}{| c |}{INFO} & 9 & 28-57 & 29-87 & 63 $\pm$ 19 & $1.4 \pm 0.2$ & $4.6 \pm 1$ \\
 \hline
\end{tabular}
\caption{Size comparison across problem classes}
\label{tab:problem_size}
\end{table*}

\subsubsection{Optimizing $|J_F|$}
\label{subsubsec:optimizing_jf}

\begin{figure*}[ht]
\includegraphics[width=0.32\linewidth]{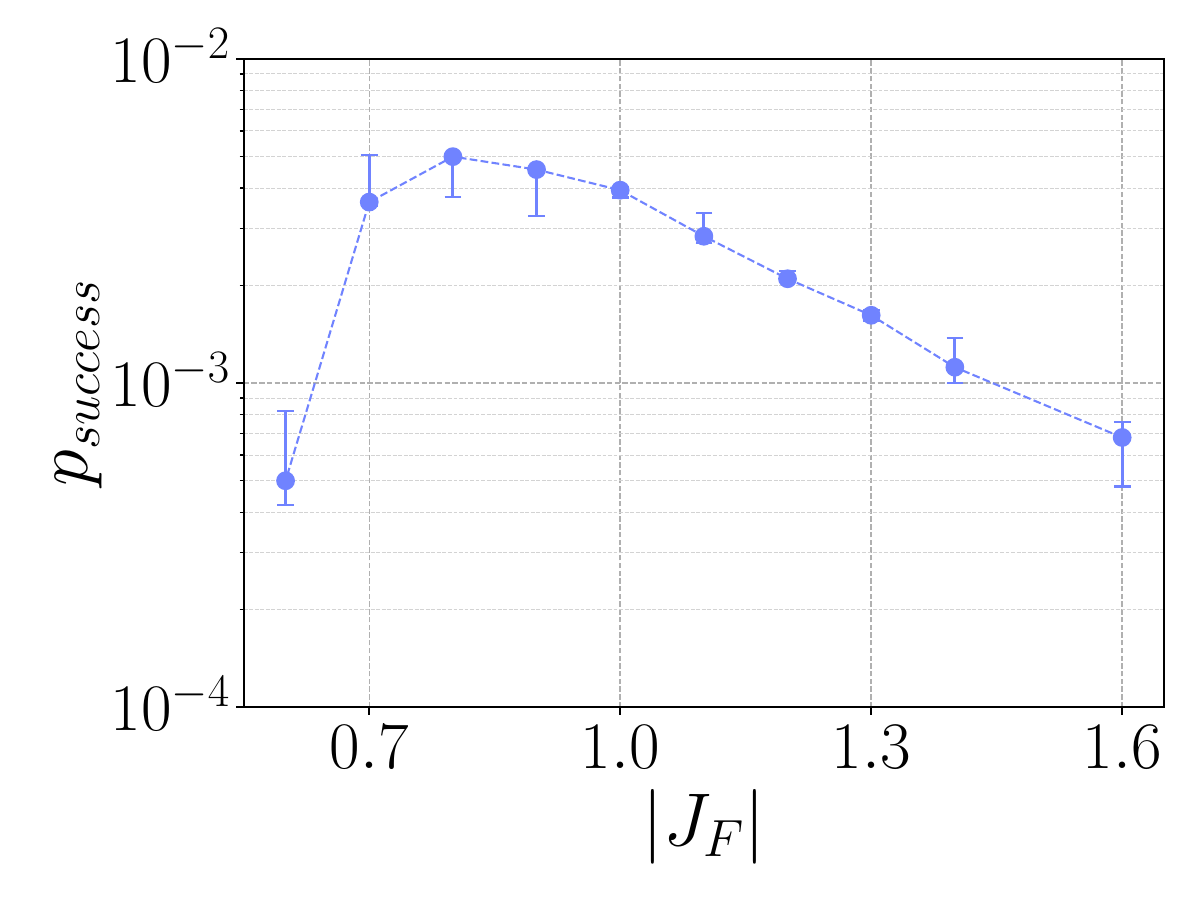}
\includegraphics[width=0.32\linewidth]{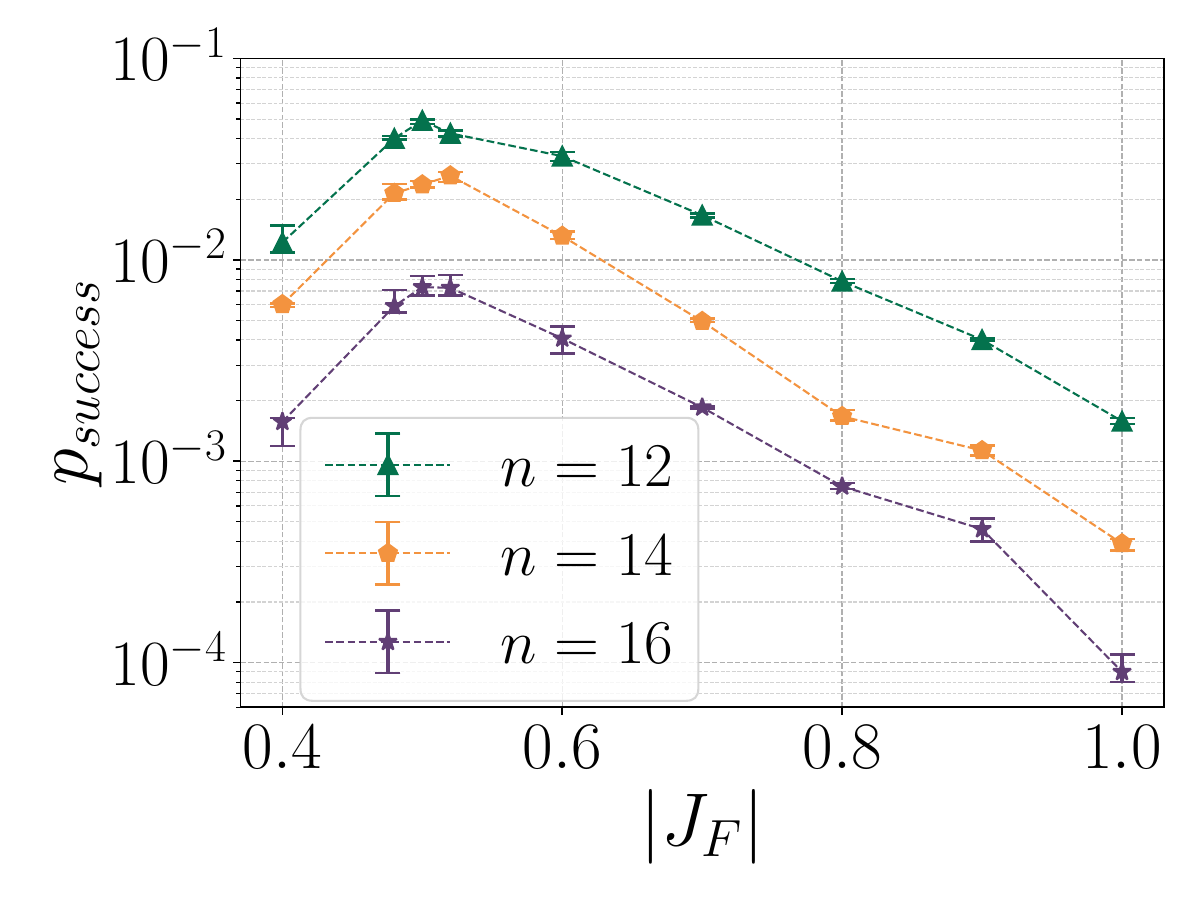}
\includegraphics[width=0.32\linewidth]{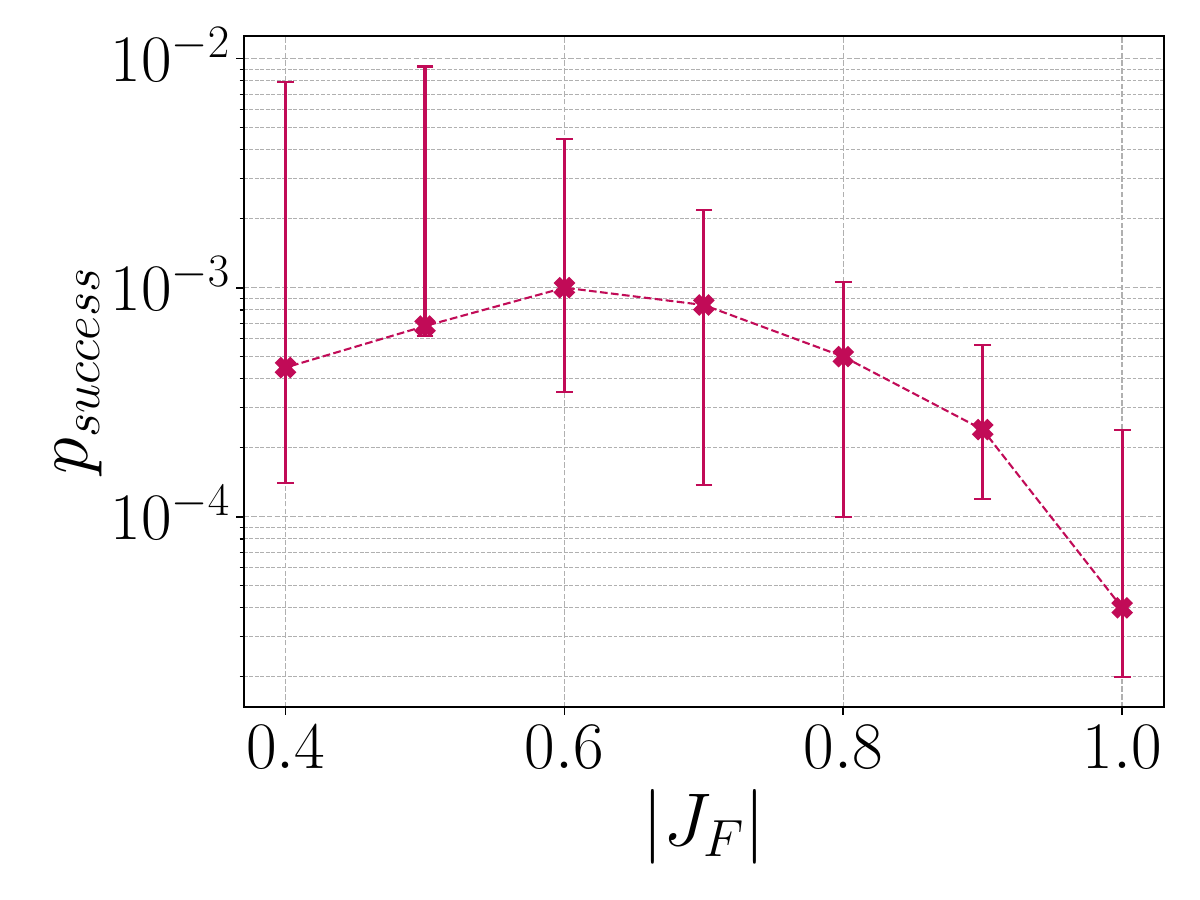}
  \caption{{\bf $p_{success}$ vs $|J_F|$ for the three different problems run on DWA.} Data points show median bootstrapped over instances, with error bars at 35th and 65th percentiles. {\bf Left:} BD MST, 45 $n=5$ instances. {\bf Middle:} GC, three ensembles of 20 instances each, with legend showing their respective $n$. {\bf Right:} INFO, nince instances.}
  \label{fig:jf_optimal_all_problems}
\end{figure*}

We start by optimizing $|J_F|$ for all three problems, as shown in Fig.~\ref{fig:jf_optimal_all_problems}. The value of $|J_F|$ influences the location of the minimum gap for the embedded problem and subsequently affects the general location of where a pause will help, as discussed in Sec.~\ref{sec:discussion}. The optimal $|J_F|$ for GC is 0.5 (0.52 for the $n=14$ instances, although not statistically different from the 0.5 result), and it is 0.6 for the preliminary results on INFO.

The optimization of $|J_F|$ presents some differences across the three problems. Let us turn our attention to the GC results first. Here, choosing the right value for $|J_F|$ provides the largest advantage, with close to 2 orders of magnitude in improvement with respect to simply setting $|J_F|=1$. Although we find $|J_F|=0.5$ to be optimal for the ensemble, there is a caveat. While these weak couplings allow configurations to change easily---the benefits of which we clearly see in the form of a much higher $p_{success}$---as the vertex model couplings get weaker, this ease of flipping also leads to a higher likelihood of the vertex model breaking, i.e. having spins that are not all aligned. When this happens, we discard the returned solutions. The fraction of solutions that contain broken vertex models increases as we decrease $|J_F|$ (in principle exponentially \cite{perils2}). A higher fraction of solutions with broken vertex models is not necessarily an undesirable feature---in fact, we find that up to a point, as that fraction increases, so does the likelihood of finding the correct minimum energy solution among the shrinking number of configurations that do not contain any broken vertex models. But there is a point of diminishing returns, when the fraction of solutions with broken vertex models is so large that the higher chances of finding the correct solution among the rest cannot compensate for their small number---this is why $p_{success}$ decreases sharply as we reach $|J_F|=0.4$. And at some point, all returned solutions contain broken vertex models, which means that the instance does not get solved, we set its $p_{success}$ to 0 and its $T_S$ to infinity. (Note that, if instead of discarding solutions with broken vertex models we chose a different approach, such as the commonly used majority vote, $p_{success}$ in these cases might not be 0, and it would likely be higher in all other cases as well.)
We find that the $|J_F|$ at which this happens stays remarkably consistent across instances, even those of different sizes. However, we do encounter some exceptions, and these can lead to one or several instances not being solved when we optimize $|J_F|$ for the majority.

In particular, of the 20 instances for each of the three sizes that are included in these results, one of the $n=12$ instances did not solve for $|J_F|=0.5$, nor did two of the $n=14$ and two of the $n=16$ instances. Of these, one of the $n=16$ solved for $|J_F| \geq 0.6$, and the other five solved for $|J_F| \geq 0.7$. Despite the unsolved instances, the median $T_S$ for the ensembles is several times smaller at $|J_F|=0.5$ than it is at $|J_F|=0.7$, when all the instances are solved.

\begin{figure}[ht]
\includegraphics[width=\linewidth]{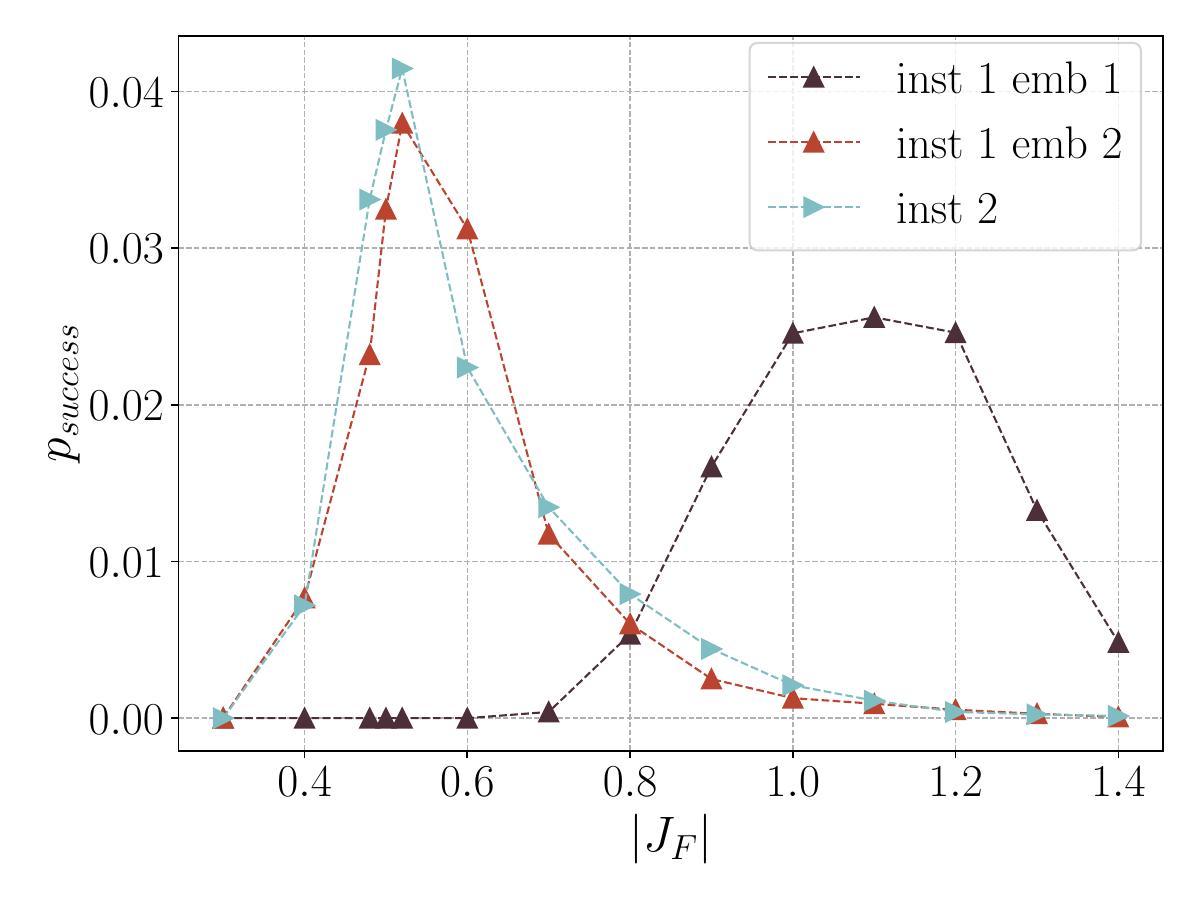}
\includegraphics[width=\linewidth]{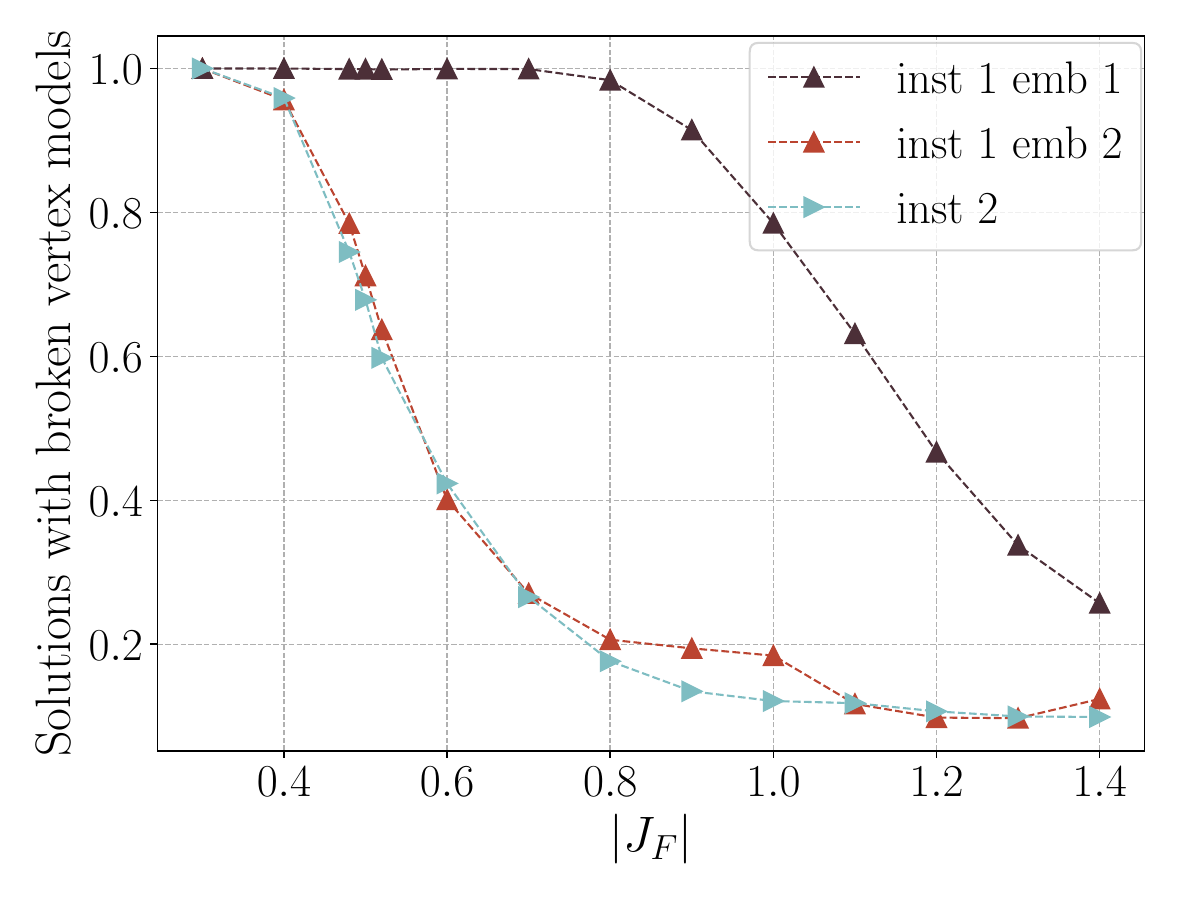}
  \caption{{\bf Optimal $|J_F|$ for two individual graph coloring instances (DWA).} 
  {\bf Top:} $p_{success}$ for the two individual $n=12$ instances, as we vary $|J_F|$.
  {\bf Bottom:} fraction of solutions returned by D-Wave where at least one vertex model was broken, and thus the solution discarded. A value of 1 indicates that all solutions found had broken vertex models.
  }
  \label{fig:2_inst_comparison}
\end{figure}

Moreover, the fact that the fraction of solutions with broken models for these particular instances reached 1 at a $|J_F|$ higher than for the rest does not appear to be related to any specific characteristics of the instances. In fact, we find that a different embedding of the same instance might show the same behavior as the majority of instances do. To illustrate this, we take the $n=12$ instance that did not solve for $|J_F|=0.5$ with its original embedding (henceforth called instance 1 embedding 1), and compare it with another $n=12$ instance of similar embedded size (instance 2), as well as to an alternative embedding to the one originally used (embedding 2). This is shown in Fig.~\ref{fig:2_inst_comparison}. The top figure shows that the original embedding used for instance 1 has a worse optimal $p_{success}$ and its location is shifted to a much larger $|J_F|$ compared with instance 2. When we select a new embedding for instance 1, however, its performance is much more similar to that of instance 2 (and of the vast majority of the instances considered). In particular, the maximum $p_{success}$ occurs at $|J_F|=0.52$ for both instances in that case. The bottom plot shows the fraction of solutions that had broken vertex models and are thus discarded. For instance 1 with the original embedding, we see that the fraction of solutions with broken vertex models saturates at a much larger $|J_F|$ than for the other two cases. This shift of the curve is similar to the one we observe in the $p_{success}$ results. The location of the optimal $p_{success}$ occurs at a similar fraction in all cases, when about $60\%$ of solutions returned have broken vertex models.

There are no obvious differences in the structure of the embedding graphs that would have led us to predict the disparities in behavior. 
Their numbers of physical qubits and edges are similar (with embedding 2 slightly higher on both counts, showcasing how choosing a smaller embedding does not always lead to better performance), and so are their mean and median vertex model sizes. It is not until we look at the embedded coefficients that we find discrepancies that can explain the different behaviors of these cases. We consider the ratio of maximum to minimum embedded coupling $R_J = \max(|J_{ij}|) / \min(|J_{ij}|)$ and that of maximum to minimum individual biases $R_h = \max(|h_i|) / \min(|h_i|)$, and find a clear difference between their values for most of the embeddings (the ones with optimal $T_S$ at or near $|J_F|=0.5$) and the few that require much stronger $|J_F|$ to be solved. For $n=12$, all embeddings have $R_J \in [15, 30]$ and $R_h \in [3, 5]$ except for the one that did not solve at $|J_F|=0.5$, which has $R_J = 7.5$ and $R_h = 2$. The second embedding for the same instance, which as described above showed a behavior consistent with the rest, had $R_J = 20$ and $R_h = 5$. Similarly for $n=14$, all embeddings have $R_J \in [15, 25]$ and $R_h \in [3.33, 5]$, except for the two that did not solve at $|J_F|=0.5$, which had $R_J = 10$ and $R_h = 2.5$. Finally, for $n=16$, all embeddings have $R_J \in [15, 35]$ and $R_h \in [3.33, 6]$, except for the two that did not solve at $|J_F|=0.5$. Of those, one had $R_J = 7.5$ and $R_h = 2.5$, while the other one had $R_J = 15$ and $R_h = 3$ (note that in the rest of the ensemble, although we find an embedding with $R_h = 3.33$ and a few with $R_J = 15$, these values do not occur together. Also, this instance solved for $|J_F|=0.6$ while the other four did not).

We can see the reason that these five instances with these particular embeddings did not solve for $|J_F|=0.5$ is that $|J_F|$ is relatively lower for them, given that the rest of the coefficients are lower. It then makes sense that the $p_{success}$ versus $|J_F|$ curve (as well as the fraction of broken chains versus $T_S$ curve) are shifted to the right compared with the rest of the instances and embeddings. Although we leave our results to include these outlier embeddings (given that we did not explore their differences until after performing our runs, and had initially decided that our method for choosing embeddings would be smallest out of 10) obtaining the $R_J$ and $R_h$ before choosing an embedding is an easy check that does not significantly increase the time resources needed to solve the problem, and can in fact give us a general idea, \emph{a priori}, of what the optimal $|J_F|$ will be.

The situation is quite different for the BD MST instances. In that case, the optimal $|J_F|$ is 0.8, but only a few instances have that value as their individual optimal. Instead, there is a relatively even spread of optimal $|J_F|$ values between 0.6 and 1.3, with the most common being 1.0 at ten occurrences. There is one instance that does not get solved at $|J_F|=0.8$, and requires $|J_F| \geq 1.1$ to find a valid solution, and we find that harder instances (i.e. those with a lower $p_{success}$, considered at their optimal $|J_F|$), tend to have higher optimal $|J_F|$. This is in contrast with the GC case, where the instance-wise optimal $|J_F|$ remains unchanged (save for the few instances discussed above) regardless of hardness.

In the INFO case, while the optimal for the ensemble is 0.6, seven out of the nine instances have their optimal at 0.5, one has it at 0.6, and the last one at 0.9 (this one is the hardest instance, in that its $p_{success}$ at optimal $|J_F|$ is the smallest out of all instances). Given that in most cases 0.5 does better than 0.6, we choose this value for our subsequent runs.  

\subsubsection{Optimizing $t_a$}
\label{subsubsec:optimizing_ta}

Setting $|J_F|$ to its optimal for each ensemble (i.e. the same for all instances), we explore a range of annealing times. Generally, the shortest annealing time is optimal in terms of $T_S$, and it is likely that these short annealing times will be particularly beneficial in combination with a pause as discussed in Sec.~\ref{sec:discussion}. When optimizing $|J_F|$, we set $t_a = 1 \mu$s, which is the shortest allowed by the annealer. We choose this value based on previous work that found it to be optimal in terms of $T_S$. We expect that increasing $t_a$ leads to higher $p_{success}$, by giving the system a longer time to thermalize. However, it is in many cases found that the increase in $p_{success}$ is not sufficient to compensate the longer time required to arrive at a solution. That is, given a set amount of time, it is more beneficial to perform many short anneals than few longer ones.

We indeed find that the optimal $t_a$ is $\leq 1 \mu$s in most cases (we cannot probe the range $t_a < 1 \mu$s so it is not possible to pinpoint the actual value). There is, however, some variation across problem classes, and nonmonotonic behavior of $T_S$ versus $t_a$, which we present in more detail in Appendix~\ref{appendix:ta_opt}. 

\begin{figure}[ht]
\includegraphics[width=\linewidth]{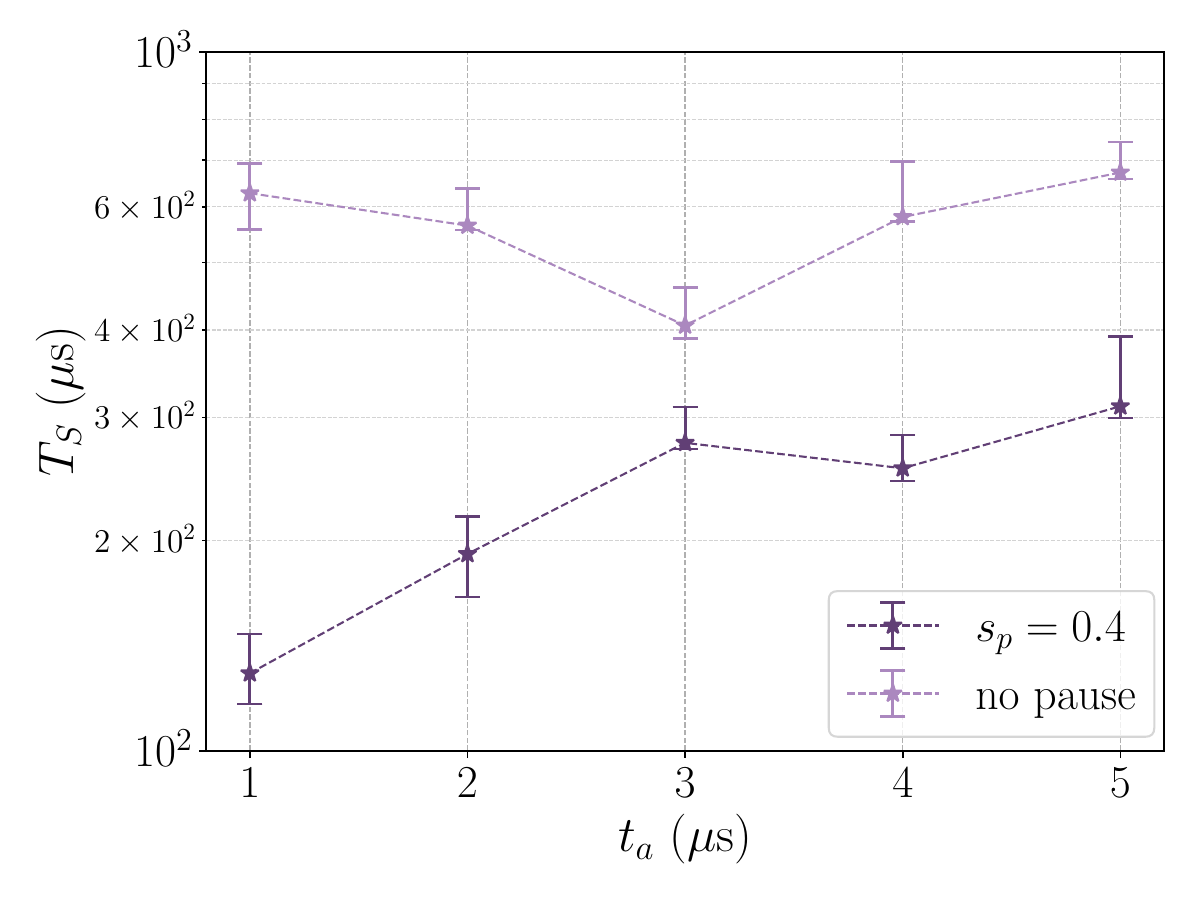}
  \caption{{\bf Optimal $t_a$ for graph coloring instances with a pause (DWA).} $T_S$ versus $t_a$ for the ensemble of $n=16$ graph coloring instances. Without a pause, the optimal in the accessible range is $t_a = 3 \mu$s. With the introduction of a 0.2 $\mu$s pause at $s_p = 0.4$, $T_S$ improves at all $t_a$, but now $t_a = 1 \mu$s does best within the range.}
  \label{fig:ta_optimal_gc_pause}
\end{figure}

One case of particular interest is the GC $n=16$ ensemble: the minimum $T_S$ in this case is found at $t_a = 3 \mu$s (note that it is possible that the true global minimum is $< 1 \mu$s). Interestingly, with the introduction of a short pause ($t_p = 0.2 \mu$s) at a good location ($s_p = 0.4$, see Fig.~\ref{fig:sp_optimal_all_problems} for results through a range of $s_p$), $t_a = 1 \mu$s does better than the nonpause minimum at $t_a = 3 \mu$s (although $T_S$ is improved by the pause for all $t_a$ tested). This is shown in Fig.~\ref{fig:ta_optimal_gc_pause}.

\subsubsection{Improving $T_S$ with a pause}
\label{subsubsec:pause_improvement}

We test whether an appropriately located pause is able to improve $T_S$. Given previous theoretical knowledge~\cite{Marshall2020,chen_pausing} and results from demonstrations~\cite{Marshall19_Pausing, gonzalez2021} about the range of locations and durations where a pause is beneficial, we explore those regions and discuss the results in Sec.~\ref{sec:discussion}. We fix $|J_F|$ and $t_a$ at the optimal found for each ensemble (in the case of $n=16$ GC instances, we choose $t_a = 1 \mu$s rather than 3 $\mu$s since we find the former to be better with a pause).

\begin{figure*}[h!t]
\includegraphics[width=0.32\linewidth]{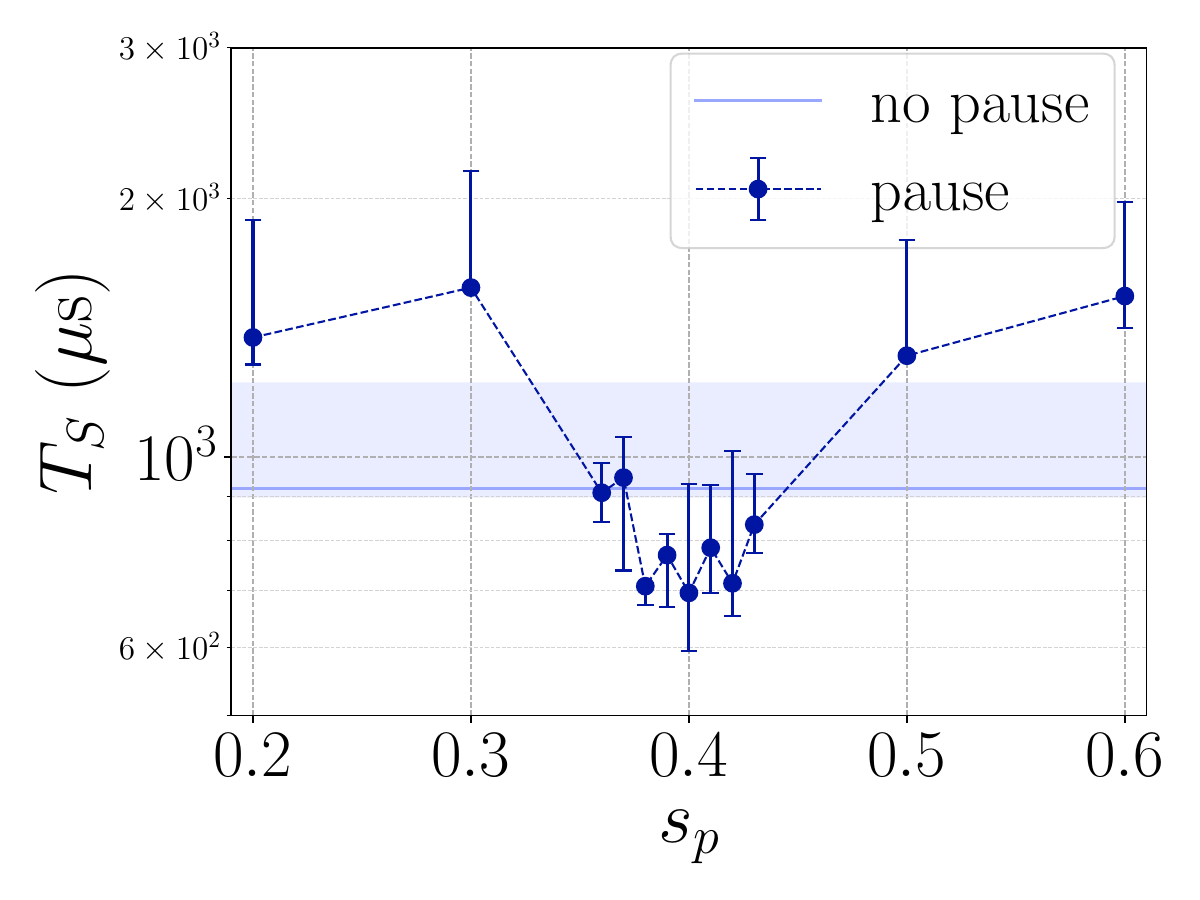}
\includegraphics[width=0.32\linewidth]{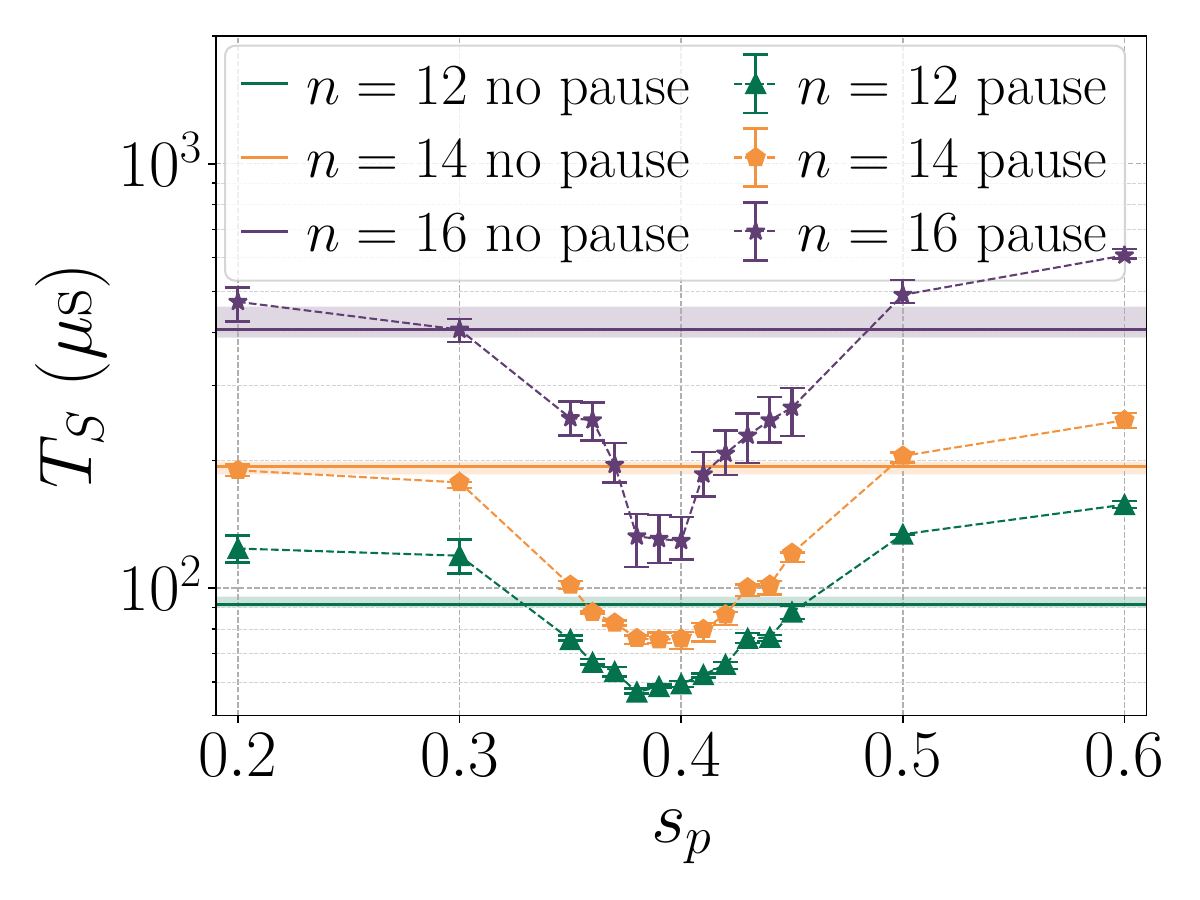}
\includegraphics[width=0.32\linewidth]{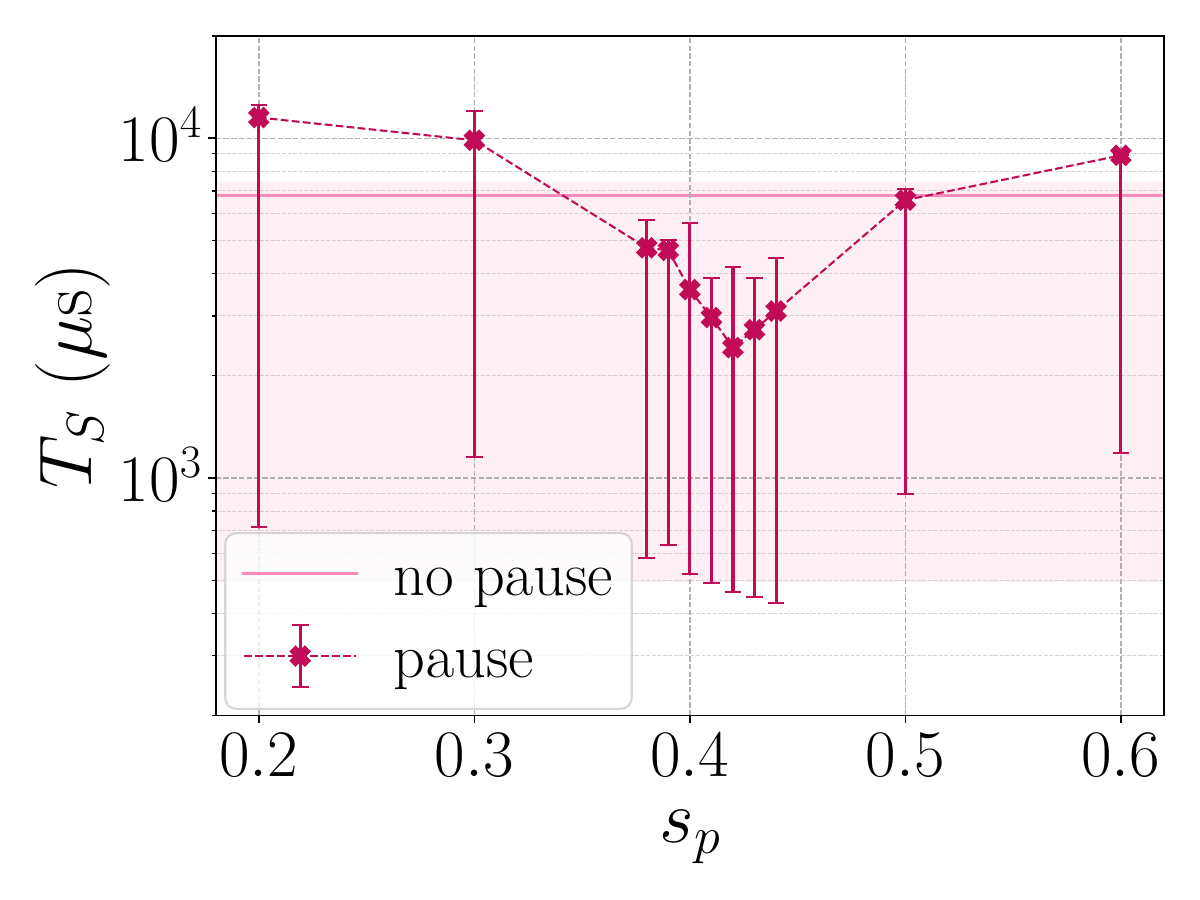}
  \caption{{\bf $T_S$ versus $s_p$ for the three different problems run on DWA.} Data points show median bootstrapped over instances for a pause of $t_p=0.2 \mu$s at the location indicated on the $x$ axis. Horizontal lines correspond to results without a pause. Error bars at 35th and 65th percentiles. {\bf Left:} BD MST, 45 $n=5$ instances. {\bf Middle:} GC, three ensembles of 20 instances each, with legend showing their respective $n$. {\bf Right:} INFO, nine instances.}
  \label{fig:sp_optimal_all_problems}
\end{figure*}

\begin{figure}[h]
\includegraphics[width=\linewidth]{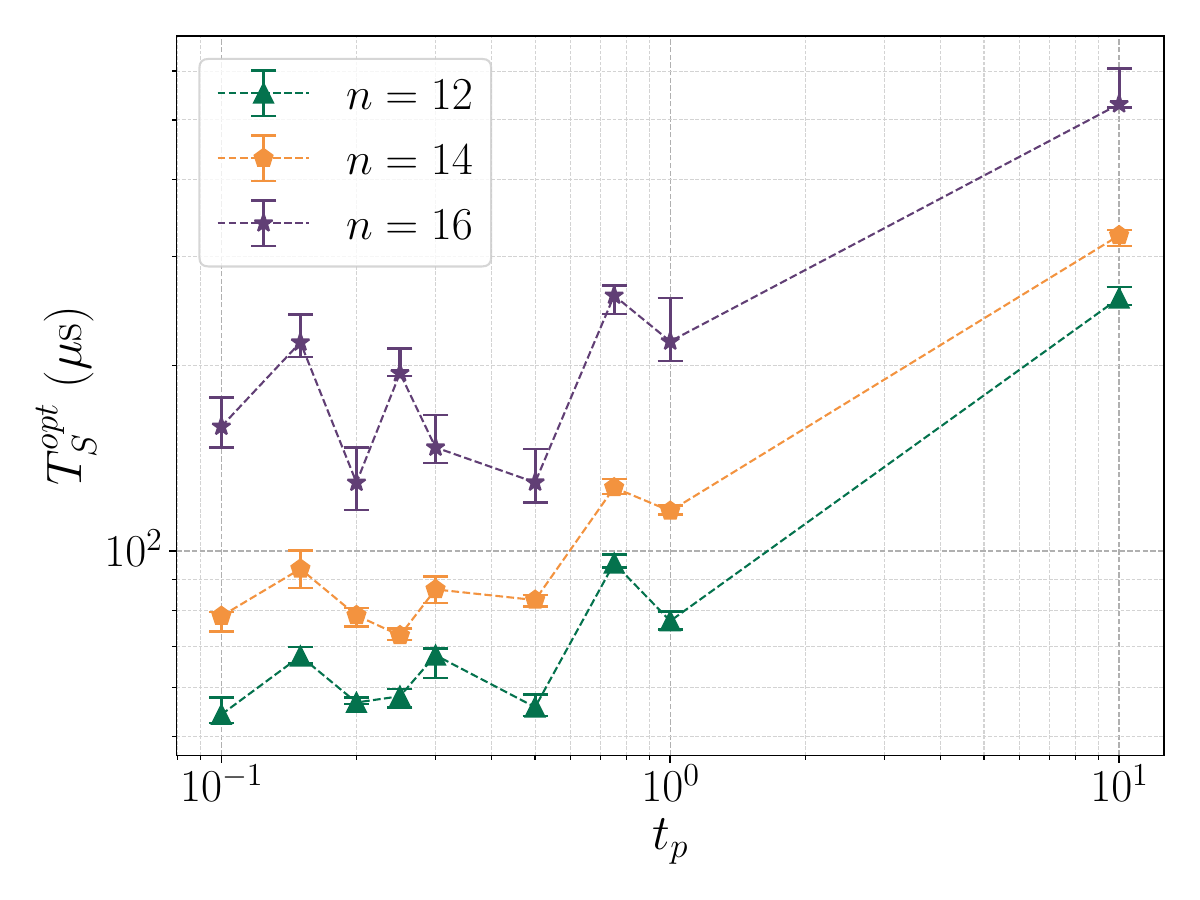}
  \caption{{\bf Best $T_S$ found for each $t_p$ (DWA).} 
  Optimal $T_S$ found for each $t_p$ for the three ensembles of 20 instances at different sizes. $|J_F|=0.5$ and $t_a=1 \mu$s always used. Note that the best $T_S$ at each $t_p$ might be found at a different $s_p$.
  }
  \label{fig:best_tts_for_tp}
\end{figure}

Fig~\ref{fig:sp_optimal_all_problems} shows $T_S$ at a range of $s_p$ between 0.2 and 0.6. We choose $t_p = 0.2$, but find no significant differences in performance in the range $t_p \leq 0.5 \mu$s for GC (shown in Fig.~\ref{fig:best_tts_for_tp}), and $t_p \leq 1 \mu$s for BD MST. INFO instances are run with $t_p = 0.2 \mu$s, 1 $\mu$s and 2 $\mu$s, and the shortest time is best, although, like all other results, not by a statistically significant amount. Given the large error bars and the results from the other two problems, we did not investigate a range of very short $t_p$. We are able to improve $T_S$ by adding a pause for all ensembles. The results for the INFO case are not statistically significant, due to having too few instances with a wide range of $T_S$. Of the nine instances, eight improve with the pause. We also repeat the runs (for the $s_p$ values that provide an improvement) for $|J_F|=0.6$, which had the best median $p_{success}$ for the ensemble, although it did worse than $|J_F|=0.5$ for eight of the nine instances. The medians for both $|J_F|$ values are comparable, with $|J_F|=0.5$ slightly better and its error bars considerably lower. 

The BD MST ensemble experiences the smallest improvement, of about 20$\%$, while the GC is closer to 40 $\%$ for $n=12$, with $n=14$ around 60 $\%$ and $n=16$ near 70 $\%$. INFO improves by almost 50 $\%$ when comparing with the $|J_F|=0.6$ results, or by around 65 $\%$ compared with those with $|J_F|=0.5$. The best pause location for the BD MST ensemble was found to be $s_p = 0.4$, although the slightly worse data point at $s_p = 0.38$ has much smaller error bars. In fact, the results are too noisy to pinpoint an exact optimal location; but we can say the region $s_p = 0.38$ - 0.42 does better than earlier or later points. Similarly, all GC ensembles perform best with a pause between 0.38 and 0.4, and the INFO between 0.4 and 0.44. The pause becomes more beneficial as these instances increase in size and hardness.

To compare the performance across different pause durations $t_p$ we first run each $t_p$ for a range of $s_p$, because the optimal location changes for different pause durations, as we discuss in Sec.~\ref{sec:discussion}. For clarity, we present these results for GC in Fig.~\ref{fig:best_tts_for_tp} by only showing the best $T_S$ found for each of the pause durations $t_p$ (on the $x$ axis), regardless of at what $s_p$ that $T_S$ was found. Our data indicates that no major differences in $T_S$ are found in the range $t_p = 0.1$-0.5$\mu$s, while once we get to $t_p = 0.75$-1$\mu$s the $T_S$ increases, and is clearly much worse as we lengthen the pause to 10 $\mu$s.

\begin{figure}[h]
\includegraphics[width=\linewidth]{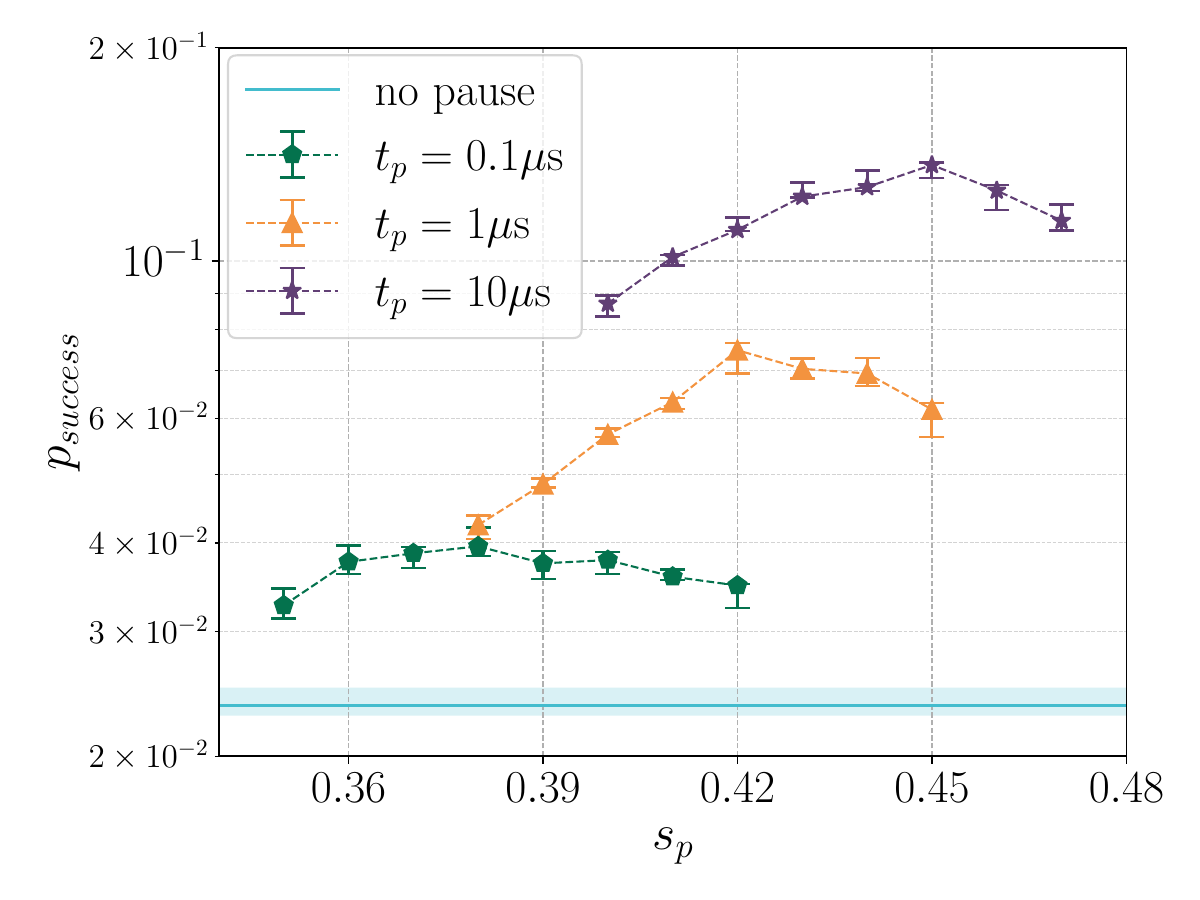}
  \caption{{\bf Shift of optimal pause location with pause duration (DWA).} 
  Peak in $p_{success}$ for three different pause durations $t_p$, for an ensemble of 60 graph coloring instances (the three ensembles of different sizes taken together). The peak shifts later as $t_p$ increases.
  }
  \label{fig:shift_sp_with_tp}
\end{figure}

As we increase the pause duration, the optimal location shifts later in the anneal. This was observed in~\cite{Marshall19_Pausing} (see Fig. 7 therein). We show the same effect, although for shorter pauses, in our Fig.~\ref{fig:shift_sp_with_tp}. In this case, the median shown in each data point is calculated over the full ensemble of 60 instances, taking all three sizes together, as is the no pause result. The peak in $p_{success}$ clearly shifts to a later $s_p$ as the pause duration increases by factors of 10. Note that although the pause durations in Ref.~\cite{Marshall19_Pausing} were longer, we maintain the relative difference between them, always a factor of 10. The shift in $s_p$ is of similar magnitude in both cases.

We note that when performing the runs to obtain these results, certain discrepancies are observed if the runs for different data points (e.g. with pauses at different $s_p$) are performed on different days. In those cases some results would appear more noisy. To avoid this, we choose to perform runs for the same $t_p$ over a range of $s_p$ one right after another for consistency. It is likely that significantly increasing the number of instances and reads would also help smooth certain results and get rid of ``noise'' due to the particulars of our ensembles, or other random factors. However, given time limitations and the additional time resources that would be required, we prefer to explore a wider range of parameters with fewer instances, this way obtaining a more general picture of parameter setting.

\section{DISCUSSION}
\label{sec:discussion}

In this section, we discuss the significance of our results and some factors that help explain them. We first review the physical picture behind pausing, and how our results agree with its predictions. Then we discuss how certain characteristics of the logical and embedded problems, such as coefficient heterogeneity and size, affect these predictions. Next, we focus more specifically on the role of these effects when optimizing $|J_F|$, followed by a discussion of optimal annealing time and some comments on the shift of the $p_{success}$ peak location with pause duration. Finally, we provide practical guidelines for parameter setting.

\subsection{Physical picture}
Let us start by reviewing the physical picture behind pausing~\cite{Marshall19_Pausing}, and the particular considerations when a pause is applied to embedded problems~\cite{gonzalez2021}. 

\begin{figure}[ht]
\includegraphics[width=0.95\linewidth]{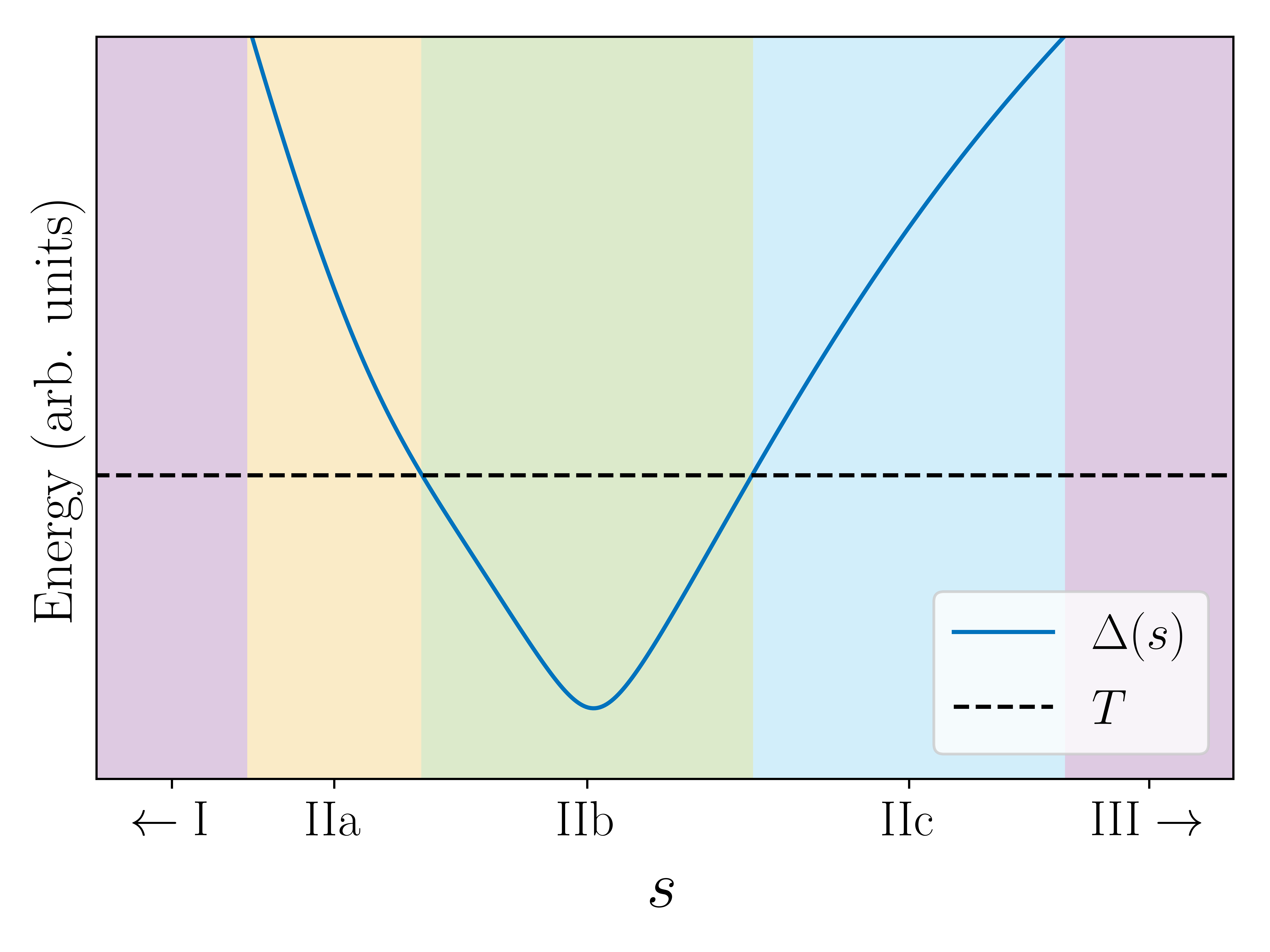}
  \caption{{\bf Diagram of the different annealing regimes.} Regimes I and III (in purple) correspond to the regions where only one Hamiltonian dominates. Regime 2, where the thermal and quantum scales are comparable, is subdivided into three regions according to the relation between the instantaneous gap $\Delta$ and the temperature $T$.}
  \label{fig:time_scales}
\end{figure}

Fig.~\ref{fig:time_scales} shows a diagram to understand the dynamics at different times in the anneal. For a more detailed explanation, the reader is directed to Sec.V of Ref.~\cite{gonzalez2021}, which we briefly summarize here: regime I corresponds to the earlier part of the anneal, when the driver Hamiltonian dominates and the system stays in its GS. Regime II encompasses the region where quantum and thermal scales are comparable. Thermal and quantum dynamics take place, both of which can transfer population from the GS to excited states. The minimum gap is in this region, and its relative size with respect to temperature distinguishes three sections with different behaviors: in section a, the gap starts out larger than the temperature and decreases approaching it. Nonadiabaticity starts to occur, although transitions (as well as thermal ones) are still slow compared with the anneal evolution. In section b, the gap is smaller than the temperature and reaches its minimum. Thermalization dominates and causes population transfer from the GS to excited states, although quantum nonadiabatic transitions also take place and contribute to the GS population loss. A long enough pause can allow the system to approach thermal equilibrium. In section c the gap is again larger than the temperature, but the quantum effects have now become weaker. The system is unlikely to follow the thermal state unless given extra time. A pause at the start of this section could help bring back some of the GS population lost to thermalization in the previous one. Finally, during regime III dynamics are essentially frozen, since the quantum scale has now become too small.

Based on this picture, we have some information \emph{a priori} about the general location in which a pause will help. We do not know, however, exactly where the minimum gap is. To further complicate matters, the minimum gap of the embedded problem will not be in the same location as that of the logical one. If instead of embedding, we increase the overall magnitude of the problem Hamiltonian $H_p$, this would amount to increasing $B(s)$ with respect to $A(s)$, hence shifting the different regions of the dynamics earlier in the anneal, including the minimum gap. A rigorous proof of this effect can be found in Ref.~\cite{Choi19}. What happens when we embed is not exactly the same; additional $|J_F|$ couplings are added to $H_p$, rather than it being simply scaled up. While no formal proof exists for this specific case that the shift of the minimum gap still occurs, we see empirically that the optimal pause location moves earlier, which points to the minimum gap also being earlier. We also provide numerical evidence of the effect for a toy problem in Sec.V.B of Ref.~\cite{gonzalez2021}.
In this scenario, two factors of the embedding would affect the shift; (1) the size of vertex models and (2) the value of $|J_F|$. Larger vertex models mean more added couplings, further adding to $H_p$, as would higher $|J_F|$ values. Moreover, these two factors typically happen in conjunction; larger vertex models are more likely to break since there are more moving parts, and thus usually require stronger $|J_F|$ to stay intact (although this is not the case under all circumstances, as we discuss later). The shift is seen in our results for the BD MST ensemble across the two devices; the optimal pause location for DW2K (when embeddings are large and $|J_F|$ strong) is found to be $s_p = 0.3$ - 0.32, much earlier than is consistently found for native problems in Ref.~\cite{Marshall19_Pausing} (around $s = 0.5$). Vertex model sizes are greatly reduced for DWA, and the optimal $|J_F|$ is much weaker. As seen in Fig.~\ref{fig:adv_vs_2kq_pause}, the optimal pausing region for DWA ($s_p = 0.38$ - 0.42) is later than for DW2K, even though the fact that $A(s)$ decays faster for DWA and crosses $B(s)$ earlier would shift it in the opposite direction. DWA's optimal pause region is still earlier than that of native problems. This all fits well with the picture of embedding shifting the minimum gap earlier.

\subsection{Problem differences: general considerations}
There are significant differences in behavior across problems, which can be related to certain characteristics of both their QUBO mappings and embeddings. Factors that determine problem hardness include size, density, and heterogeneity of coefficients, i.e., how many different values of $J_{ij}$ and $h_i$ are present, and how close those values are to one another. The devices we use suffer a range of integrated control errors (ICEs)~\cite{dwave_ice}, due to which the coefficients implemented in practice will differ from the user inputted ones by some $\delta J_{ij}$ and $\delta h_i$. 
Precision is in this way limited, and having too many different coefficients or some that are too close together can result in them getting mixed up once we account for $\delta J_{ij}$ and $\delta h_i$. The number of different coefficients increases after embedding and their values change, as they get split over several qubits and couplers. Large vertex models are detrimental since they can lead to too small coefficients, and so is having several vertex model sizes, which add more different coefficients, potentially closer to one another. In the standard embedding algorithm included in the D-Wave software, the $h_i$ corresponding to a single variable is evenly distributed over the physical qubits of the vertex model (between one and six for all the problems presented in the main text, although the largest instances of $n=6$ BD MST in Appendix~\ref{appendix:mst_n6} have maximum vertex model size going up to 10), and the $J_{ij}$ for a logical coupling also gets evenly split over all the available physical couplers between qubits in each vertex model (between one and eight couplers for the problems in the main text, going up to nine for the largest $n=6$ BD MST instances).

A more dense problem is more likely to require a larger embedding, and larger vertex models will be needed to accommodate the extra connectivity. The more dense embedded problem will lead to frustrations and a higher rate of broken vertex models and discarded solutions. Larger and denser problems are in principle more difficult, but in our demonstrations we find this secondary to the characteristics of the coefficients. This becomes clear when we consider the differences in $p_{success}$ or $T_S$ across ensembles. For example, as we can see in Fig.~\ref{fig:jf_optimal_all_problems}, the three GC ensembles have $p_{success}$ highest for the smallest ensemble ($n=12$), subsequently decreasing for the $n=14$ and $n=16$ cases. However, all three GC ensembles have higher $p_{success}$ than the BD MST and INFO ones, even though they are the largest in size. The embedded size of the smallest GC instances is similar to the larger BD MST ones, and the $n=14$ and $n=16$ upper range goes well beyond that of the BD MST. This is even more pronounced for the INFO case, which has the smallest instances, and yet its $p_{success}$ is the lowest of the three problems. GC instances are typically denser than BD MST ones, with average degree $5.1 \pm 0.3$ and little variability for different $n$, while BD MST has average degree $4.6 \pm 0.3$. INFO instances have average degree $4.6 \pm 1$.

\subsection{Coefficient heterogeneity, hardness and optimal ferromagnetic coupling}
Based on our results, we present the hypothesis that the main differentiating feature across problem classes is coefficient heterogeneity. The coefficients of GC instances are remarkably regular, and regardless of the problem specifics they will not have very disparate values. The GC QUBO mapping (Appendix~\ref{appendix:mapping_gc}) does not include an objective function, but only two penalty terms, which means that, unlike the other problems, it does not require a penalty weight factor acting on the penalty terms to ensure that violating one does not become advantageous. Thus, all coefficients of the GC QUBO are either one or two. Once embedded, there is a wider range of values as the $h_i$ and $J_{ij}$ get divided and assigned by the embedding heuristic, which leads to a small set of coefficients that the majority of instances share. We do find a discrepancy, discussed in Sec.~\ref{subsubsec:optimizing_jf} between the ~93$\%$ of instances with optimal $|J_F|=0.5$ and the rest with a different optimal $|J_F|$; this difference can be traced to the fact that the minimum vertex model size in the majority is one, while in the cases with a higher optimal $|J_F|$ it is two. Yet, the physical qubits in those smallest vertex models are assigned the same $h_i$ value whether there is a single physical qubit or two, resulting in the sum of all the biases corresponding to a given vertex model, or all the physical couplings representing a single logical one to be twice as large in the few instances that behave differently, thus needing a roughly twice as strong $|J_F|$ to obtain the best performance. This is a detail that one should keep in mind when using this embedding routine without further examination, as it can lead to unexpected behavior like in this case.

The level of coefficient heterogeneity can explain the greater impact that optimizing $|J_F|$ has on the GC ensembles compared with the other problems. All the individual curves (save for the few exceptions) look the same as that of the ensemble. On the other hand, for BD MST we have many different individual curves, with maxima across a range of $|J_F|$ values, resulting in a much less sharp peak when averaged. For INFO, although the optimal is fairly consistent, the small number of instances and wide range of $p_{success}$ lead to very large error bars, somewhat diluting the results.

Compare the QUBO formulation for GC with those of BD MST or INFO. In the BD MST QUBO (Appendix~\ref{appendix:mapping_mst}), we have a cost function that has each of the different weights as coefficients. Then, the penalty terms can have coefficients 1, 2, and 4, multiplied by the penalty weight, which needs to be larger than the maximum weight. Before embedding, we are already at quite a few different values, depending on the number of weights. For this reason, we avoid instances with weights that are too disparate, which will quickly run into precision issues. Intuitively, those instances would seem easier to solve, since it would be our first instinct to avoid the more costly parts of the graph, while for the annealer, these are more difficult. For INFO (Appendix~\ref{appendix:mapping_info}), this can become even more pronounced. The coefficients for the cost function run through the product of each individual cost with the possible times at which the message will be traveling, so we have two parameters coming into play. This can result in a wide array of values, which we curb by setting a short time horizon and keeping all other problem parameters within narrow ranges (e.g. two-four messages, delay costs of one-three, travel time for each edge one-two). Despite this similarity among instances, and their yielding the smallest embedded sizes (compared with the other two problems), their median difficulty is the highest out of the three ensembles, and their disparity even larger, including the instance with the highest $p_{success}$ as well as some of the most difficult ones. The penalty terms coefficients can be one or two, and are also multiplied by a penalty weight, which depends on costs and times as described in Appendix~\ref{appendix:mapping_info}. Just like for BD MST, costs that are too disparate can take us beyond precision (and so do long times, which can be a result of long travel times in the connectivity graph, having long paths, or not setting a tight time horizon bound), leading to problems that are difficult for the annealer but would be intuitively easy to solve, by not delaying the message(s) with the much higher cost(s) than the rest.

This picture correlates well with what we see in our instances. GC instances have (once embedded) between six and ten different $J_{ij}$ values, and three to six different $h_i$ ones. INFO instances have 4-11 for $J_{ij}$ and 11-30 for $h_i$, and BD MST instances have 7-14 for $J_{ij}$ and 12-27 for $h_i$.

\subsection{Relationship between vertex model size and optimal ferromagnetic coupling}
Our results on the differences in optimal $|J_F|$ challenge the conventional wisdom that smaller vertex models are less likely to break and thus will do better with a weaker $|J_F|$. While this is the correct approach in some cases (e.g., solving the BD MST ensemble in DWA had smaller vertex models and a much lower optimal $|J_F|$ compared with DW2K), things are more subtle when other differences are present. We see how most of the GC instances have a clear optimal at $|J_F|=0.5$, while BD MST instances have a range of optimal values between 0.6 and 1.3. Yet, the average vertex model size for GC instances is $2.8 \pm 0.2$ for both $n=12$ and 14, and $3.0 \pm 0.2$ for $n=16$, while for BD MST it is significantly smaller at $1.7 \pm 0.3$. We hypothesize that the characteristics of the rest of the coefficients seem to have a greater impact on optimal $|J_F|$ than simply the size of the vertex models. Density could also play a complicated role. Similar to what happens with size, a denser vertex model has more competing forces that can lead to breaking, which would point towards a higher $|J_F|$ being needed to keep it consistent. Yet, that density could also require more ease to change configurations among the increased number of possible ones, thus benefiting from weaker $|J_F|$. We see in Fig.~\ref{fig:2_inst_comparison} that the optimal $p_{success}$ is reached when ~60$\%$ of solutions returned are broken, and that number stays consistent between the two different embeddings, despite their disparate coefficients. When considering instances of the same BD MST ensemble, where there are some commonalities among coefficients, although not to the degree of the GC case, the smaller vertex models with weaker $|J_F|$ correlation does apply. Within the BD MST ensemble, we find that larger instance size correlates with larger vertex models, stronger optimal $|J_F|$, and higher $T_S$.

A result that embodies these last two points is that the optimal $|J_F|$ remains constant through changes in hardness for GC problems, while for the BD MST instances it increases with hardness. This points to the fact that, when hardness comes mostly from overall size, it does not affect optimal $|J_F|$, while when it is due to the other factors a stronger $|J_F|$ is required to keep up.

A more quantitative analysis to tease out the effects of each of these aspects will be necessary to inform parameter setting across different problems and platforms, which becomes critical as difficulty increases, as choosing the right parameters can determine whether a valid solution is found at all.

\subsection{Optimal annealing time}
We see in Sec.~\ref{subsubsec:optimizing_ta} that in most cases the shortest annealing time is optimal in terms of $T_S$. The exception is the GC $n=16$ ensemble, which clearly performs better at slightly longer $t_a$. These instances have the largest sizes out of all our ensembles. Although D-Wave quantum annealers can in principle be operated with shorter than 1 $\mu$s times, this capability has not yet been offered to the public (but it has recently been announced that shorter times of up to 0.5 $\mu$s will be made available in the near future). It is likely that these shorter times would be particularly beneficial in combination with a pause. When we introduce a short pause of 0.2 $\mu$s at an appropriate location for the GC $n=16$ ensemble, the optimal $t_a$ is no longer 3 $\mu$s like in the no pause case, but instead $\leq$ 1 $\mu$s like for the other ensembles, as shown in Fig.~\ref{fig:ta_optimal_gc_pause}. This tells us that 0.2 $\mu$s of extra time at a well-chosen location is more beneficial than up to 9 $\mu$s of extra time spread equally over the whole anneal (since in the no pause case, while the optimal is at 3 $\mu$s, several $t_a$ up to 10 $\mu$s had lower $T_S$ than 1 $\mu$s). This is also seen in Fig.~\ref{fig:sp_optimal_all_problems}, where a pause too early or too late leads to a higher $T_S$. There are regions in the anneal where going faster is unlikely to have a detrimental effect, such as when dynamics become very slow, and that time can be better used elsewhere to improve $T_S$. It is also unclear whether we have find the true optimal $t_a$ at 1 $\mu$s for our ensembles, or lowering $t_a$ would further decrease $T_S$. Access to a lower range of $t_a$ will help answer these questions.

\subsection{Peak shift with pause duration}
We also verify, for the ensemble of GC instances across all three sizes, the $p_{success}$ peak shift with $t_p$ observed for native problems in Ref.~\cite{Marshall19_Pausing} (see Fig.~\ref{fig:shift_sp_with_tp}). The reason for this shift is that, to maximize our chances of obtaining the GS of the problem Hamiltonian $H_p$, it would be advantageous to have the system thermalize as late as possible in the anneal; ideally at $s=1$ when the instantaneous Hamiltonian is $H_p$, so the instantaneous GS is that of $H_p$. In practice, the system is not able to thermalize too late in the anneal, after the freeze-out point~\cite{Amin15, Marshall17}, and even before reaching this point it becomes progressively more difficult as quantum fluctuations decrease. Introducing a pause is a way of making thermalization more likely to happen. As we move later in the anneal, a longer pause is required to give the system enough time to thermalize, but because the instantaneous GS is becoming closer to that of $H_p$, the benefit is larger. This is why we also see $p_{success}$ increasing as the peak shifts later with the lengthening of $t_p$. However, as we have seen in Fig.~\ref{fig:best_tts_for_tp}, that increase in $p_{success}$ is not sufficient to compensate the additional time, and $T_S$ worsens for longer pauses.

\subsection{Practical recommendations}
We devote this section to examining how the different annealing parameters relate to one another, depend on problem characteristics and affect $T_S$. But a reader hoping to attack a new problem on a quantum annealer might still be left wondering what parameters to choose, especially if time is a constraint and exploring a range of parameters not possible. Here, we take all our findings into consideration to provide a set of guidelines that can be applied in practice. Given that problems vary widely, these guidelines should be understood as a starting point and a means to obtaining good results without the overhead of benchmarking, but cannot be expected to provide optimal performance without additional tweaking and exploration.

Let us start with setting $|J_F|$. This parameter has the largest effect on $p_{success}$ and does not increase running time, so if additional resources are available, it is a good idea to use them here and test more than one value. First, consider the QUBO formulations of the set of instances to be solved. Do they all have very similar coefficients, or is there a lot of variety? If the former is true, they will likely all have the same (or very similar) optimal $|J_F|$, which simplifies things. The best course of action is then to explore a range of values for a single instance, identifying its optimal value, and running the rest with that value only.

If, on the other hand, the coefficients vary a lot across instances, optimal $|J_F|$ will too. In this case, one must decide how much extra work to put into choosing different $|J_F|$ for each instance. Assigning the same to all is the simplest and quickest way, but can lead to poor results for those at the tail ends, in particular those with very high optimal $|J_F|$, for which a value in the middle can mean that no valid solutions are returned due to all of them having broken vertex models. In this case, because the strength of the optimal $|J_F|$ correlates with problem size, vertex model size and number of different coefficients, it can be worth splitting the ensemble into subgroups according to one of those quantities and assigning a different $|J_F|$ to each of them (logical size is a good straightforward choice, although if there are very large differences in vertex degree across instances, embedded size will be a better predictor). A range between 0.5 and 1.5 can be a good start. Using a majority vote approach to process solutions with broken vertex models, rather than discarding them like we do here can also be helpful.

It is not possible to determine whether $|J_F|$ is close to its optimum by looking at the results for a single value. We can, however, get a rough idea of whether we are in the right region by looking at the fraction of solutions with broken vertex models, which is a simple check. A majority of the returned solutions will contain broken vertex models when performance is best, but not such a large fraction that obtaining valid solutions becomes almost impossible. After testing a $|J_F|$ value, if the fraction is close to 100$\%$, a stronger $|J_F|$ should be used. If it is small, a weaker $|J_F|$ is necessary. 

For the annealing time $t_a$, we have seen that keeping it at the current shortest possible (1 $\mu$s) combined with a pause yields the best results. We suspect that this will hold even as shorter times become available, but there will be a (currently unknown) limit where results worsen with shorter times, so one should be careful if $t_a \ll 1 \mu$s becomes available. The pause duration $t_p$ should also be short. We have not found measurable differences in the range $t_p \leq 0.5 \mu$s, and consider $t_p = 0.2 \mu$s a safe bet.

A pause location $s_p$ within the region 0.3-0.5 will generally improve performance. The smaller the vertex models, the later the optimal pause location and vice versa. For instance, for an embedding that requires only a few extra qubits, and where many of the logical variables remain unembedded, a pause close to 0.5 will likely be best. For the instances in this paper, where vertex models are typically one-five qubits (when using DWA), we find the optimal location around 0.4. The results for the older DW2K, with larger vertex models, did best closer to 0.3. For really large vertex models this location will probably have to be pushed even earlier (but we do not currently have the empirical data to confirm it). If the $A(s)$ and $B(s)$ functions are significantly different from the ones used in this work, and in particular the region where their scales are comparable is shifted, the pause location might need to be adjusted in the same direction as this shift.

Finally, one should keep in mind that time and other resources required to find optimal (or even simply good) parameters must be taken into account. When we report $T_S$ for a set of parameter values, the time needed to find them is not included. But that time will be a limiting factor for practical applications, which is considered to provide the above guidelines.

\subsection{A note on our optimization strategy}
In our benchmarking study, we mainly follow the strategy of optimizing a single parameter at a time while keeping the rest fixed, and thus there are large regions of the multidimensional parameter space that have been left unexplored. Although this means that we cannot provide empirical evidence that the optimal set of parameters resides in the region that was investigated, we believe this to be the case, and our method to be a reasonable choice. First, it would be not just impractical, but virtually impossible to obtain results covering the entirety of this space, so making some assumptions is unavoidable. With this in mind, we base our runs on the physical picture described at the beginning of this section, whose predictions have matched our observations well. We then make adjustments based on the trends that the first rounds of runs revealed and, when possible, cover a large range of values to ensure the observed trends are not local to a small region. Finally, in the cases where there is a physical reason for a parameter to affect another optimal value, this is taken into account---for instance, after finding an optimal $t_a > 1 \mu$s for certain instances with an initial no pause schedule, several $t_a$ values are explored again after the introduction of a pause.

\section{CONCLUSIONS}
\label{sec:conslusions}

We investigate the physical picture of pausing on quantum annealers, in combination with the parameter setting problem, through demonstrations across multiple devices and non-native problems. We gained insights regarding the characteristics of the logical problem---and their transformation through embedding---that impact the hardness of the problems, as well as the parameters that optimize their time to solution $T_S$.

Pausing midanneal was first proposed as a strategy to improve the probability of success $p_{success}$ for native optimization problems~\cite{Marshall19_Pausing}, and was later confirmed to work for embedded problems~\cite{gonzalez2021}, and to also improve $T_S$ with a careful choice of pause duration~\cite{Albash2021}. We study an ensemble of problem instances on two quantum annealing devices with different architectures and annealing schedules, and also three ensembles of different problems on a single device, leading to further confirmation that pausing can improve $T_S$ in a variety of scenarios. We find the region of improvement stays fairly robust, barring small adjustments due to large differences in vertex model size or its corresponding ferromagnetic coupling $|J_F|$. Smaller vertex models and smaller $|J_F|$ shift the location of the region later, and vice versa.

We provide evidence that adding a short pause at a well-chosen location boosts $T_S$ more than a much longer annealing time overall does. Combined with the fact that a pause outside of the beneficial region worsens $T_S$, this gives empirical support to the idea that there exist rather large sections at both ends of the anneal where annealing faster will not hinder performance (up to a point that cannot currently be determined due to hardware limitations), and that pauses will likely improve $T_S$ even when the optimal annealing time $t_a$ is reached.

Along with the pause location $s_p$, we also find the optimal annealing time $t_a$ and pause duration $t_p$ to remain consistent across the different scenarios, while the ferromagnetic coupling within vertex models, $|J_F|$, requires careful examination to be chosen. Nonetheless, optimizing $|J_F|$ had a much larger effect on performance than any of the other parameters (potentially orders of magnitude greater). 

We observe a correlation between coefficient heterogeneity (having many different values for $J_{ij}$ and $h_i$) and problem hardness across different problem classes, while size played a bigger role within a given class. We also saw that when coefficient heterogeneity is low, the optimal $|J_F|$ stays constant through changes in size (and hardness), while in the opposite case (high heterogeneity), larger instance size correlated with larger vertex models and stronger optimal $|J_F|$. The conventional wisdom that large vertex models require stronger $|J_F|$ only held up within the ensemble with high coefficient heterogeneity, while it proved incorrect when applied across different problem ensembles as well as when coefficient heterogeneity is low within an ensemble.

A quantitative analysis of the effects that these characteristics have on hardness and parameter setting, and their interplay, should be a fruitful field for future study and pave the way for more rigorous parameter setting. The same can be said of the relationship between optimal pause location and vertex model size. 
As new devices with more connected architectures become available, embedding strategies such as utilizing all the available couplers between vertex models might need to be reconsidered. These devices will also support shorter annealing times which will allow us to verify what we already strongly suspect---that additional time is only useful in certain regions, so a pause lowers optimal annealing time and will be able to improve performance even when that optimal $t_a$ is achievable. Theoretical and empirical limits on how short is too short will require investigation, as will further tailoring of the annealing schedule, such as optimizing annealing speed at different parts of the anneal.

\section{Acknowledgements}

We are grateful for support from NASA Ames Research Center, the Autonomous Systems discipline in the Transformational Tools and Technologies (TTT) Project of the NASA Transformative Aeronautics Concepts Program, and from DARPA under IAA 8839 Annexes 125 and 128. 
Z.G.I., J.M., and Z.W. are thankful for support from NASA Academic Mission Services, Contract No. NNA16BD14C. We thank Alexander Sadovsky for helpful feedback on the information sharing problem. Z.G.I. thanks David Bernal Neira for useful discussions.

\bibliography{main}

\appendix

\newpage

\section{Quantum annealing devices}
\label{appendix:device_details}

The two devices we use are D-Wave 2000Q (DW2K), and D-Wave Advantage (DWA). We discuss some of their features relevant to our study below, but more in-depth technical information can be found in Ref.~\cite{DWave} for DW2K, and in Ref.~\cite{advantage_report} for DWA.

Let us first talk about the similarities and differences between the two quantum annealers. Both devices perform qualitatively the same process, implementing the time-dependent Hamiltonian
\begin{equation}
H(s) = A(s)\sum_i \sigma_i^x + B(s) \left( \sum_{<i,j>} J_{ij} \sigma_i^z \sigma_j^z + \sum_i h_i \sigma_i^z \right),
\end{equation}
with $s$ the dimensionless time parameter going from 0 to 1 to perform one anneal. Their driver Hamiltonian is the same, a transverse field over all qubits that produces quantum fluctuations. Other quantitative details, however, differ.

\begin{figure}[h]
\includegraphics[width=\linewidth]{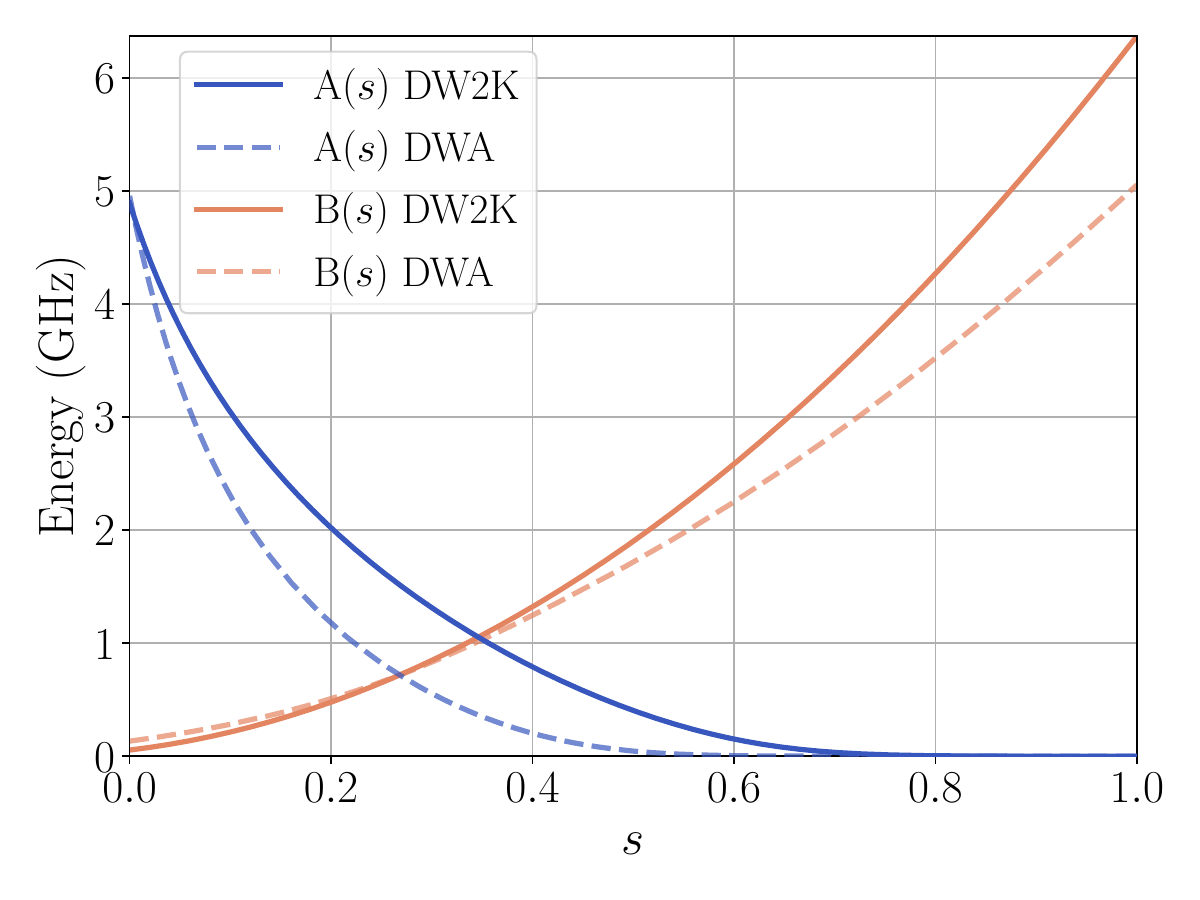}
\includegraphics[width=\linewidth]{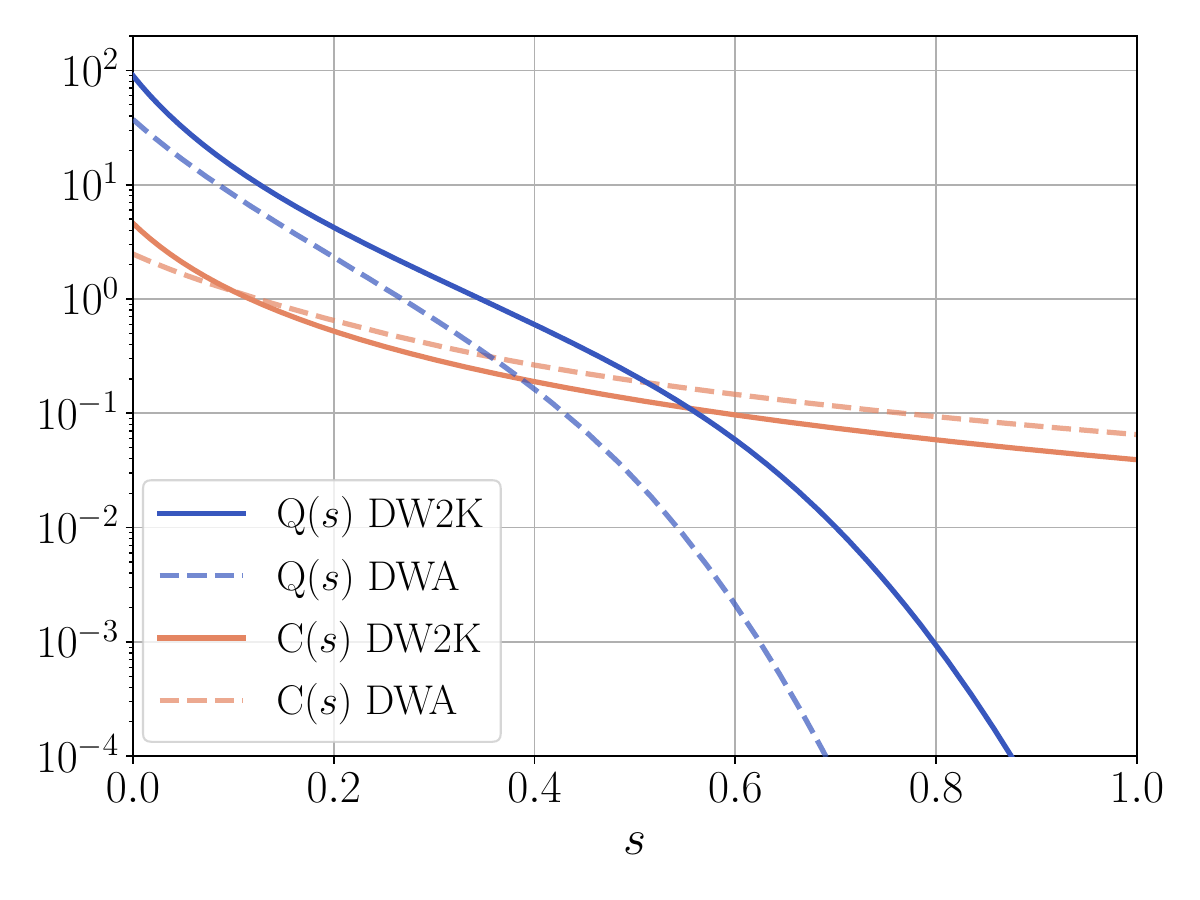}
  \caption{{\bf Top: $A(s)$ and $B(s)$ functions for the two devices.} 
  Energy scales for the driver and problem Hamiltonian of the two devices we use in our demonstration, in units of GHz, and $h=1$.
  {\bf Bottom: $Q(s)$ and $C(s)$ for the two devices.} 
  Dimensionless quantum ($Q(s)$) and classical ($C(s)$) scales for the two devices we use in our demonstration.}
  \label{fig:ab_functions}
\end{figure}

The annealing functions $A(s)$ and $B(s)$ present somewhat different shapes, as shown in Fig.~\ref{fig:ab_functions} (top). The $B(s)$ is fairly similar in both devices until the later part of the anneal, where DW2K ends up higher, and $A(s)$ decays earlier and more rapidly for DWA than for DW2K. Their relative strength is what determines the different regions in the dynamics of the system, so it is more convenient to look at two dimensionless scales as defined in Ref.~\cite{Marshall17}; $Q(s) = A(s) / B(s)$ corresponding to quantum fluctuations and $C(s) = k_B T / B(s)$ to thermal ones. Note that the temperature of the devices is different, with DW2K running at $T=12.1$ mK and DWA at $T=15.8$ mK. These scales delineate three regimes within the anneal: when the quantum fluctuations are much stronger than the thermal ones early on, the system closely follows the GS of $H(s)$; in the middle, where both scales are comparable, most of the dynamics takes place, and towards the end, once the quantum fluctuations become much smaller than the thermal ones the dynamics are essentially frozen, and no more transitions occur. But as we can see in Fig.~\ref{fig:ab_functions} (bottom), while the thermal fluctuations follow a similar curve in both devices, the same is not true for the quantum ones. $C(s)$ and $Q(s)$ cross much earlier in the anneal for DWA, meaning the region of stronger dynamics moves earlier.

The most obvious and significant difference between the two devices is in their number of qubits and their connectivity. DW2K has a 2048-qubit architecture (with 2031 working qubits), arranged in a Chimera graph. See the top panel of Fig.~\ref{fig:architectures} in Appendix~\ref{appendix:device_details} for a representation. Bipartite cells with eight qubits each are arranged in a square pattern, connected to those on either side as well as above and below. Except for those on the edges, each qubit is connected to six others. A limitation of this architecture is its inability to natively support odd cycles. DWA has a much larger and connected architecture, with a 5760-qubit graph (5436 working ones) arranged according to a Pegasus graph~\cite{pegasus_report}, as shown in the bottom panel of Fig.~\ref{fig:architectures}. This graph includes Chimera as a subgraph, but adds an extra layer of connectivity, bringing each nonedge qubit up to degree 15, and making odd cycles available.

Although decreases in noise and improvement in fabrication have an impact on the quality of results~\cite{Pokharel2021}, we expect that the largest difference in performance will be due to the reduced embedding size. The fabrication process has not changed between the two devices we are using. (There is a low-noise version of DW2K which used a different fabrication process, but we did not have access to it~\cite{DWave_low_noise}). The quality of solutions depends in part on implementing the correct problem. This is determined by integrated control errors~\cite{dwave_ice}, which encompass several sources of problem infidelity. How much the actual values of couplings $J_{ij}$ and fields $h_i$ will differ from the programmed ones depends on many factors, including the specific values being programmed, the annealing parameter $s$ and the time at which the problem is run, so it is difficult to determine how much these errors are affecting our results. Measurements for most of the error sources have not yet been released for DWA~\cite{advantage_qpu} so, given our current knowledge, we cannot presume any significant differences in the effect of ICE for both devices.

Fig.~\ref{fig:ab_functions} shows the annealing functions $A(s)$ and $B(s)$ for both devices used in our demonstration. Fig.~\ref{fig:architectures} shows the graphs of their respective architectures.

\begin{figure}[h]
\includegraphics[width=0.8\linewidth]{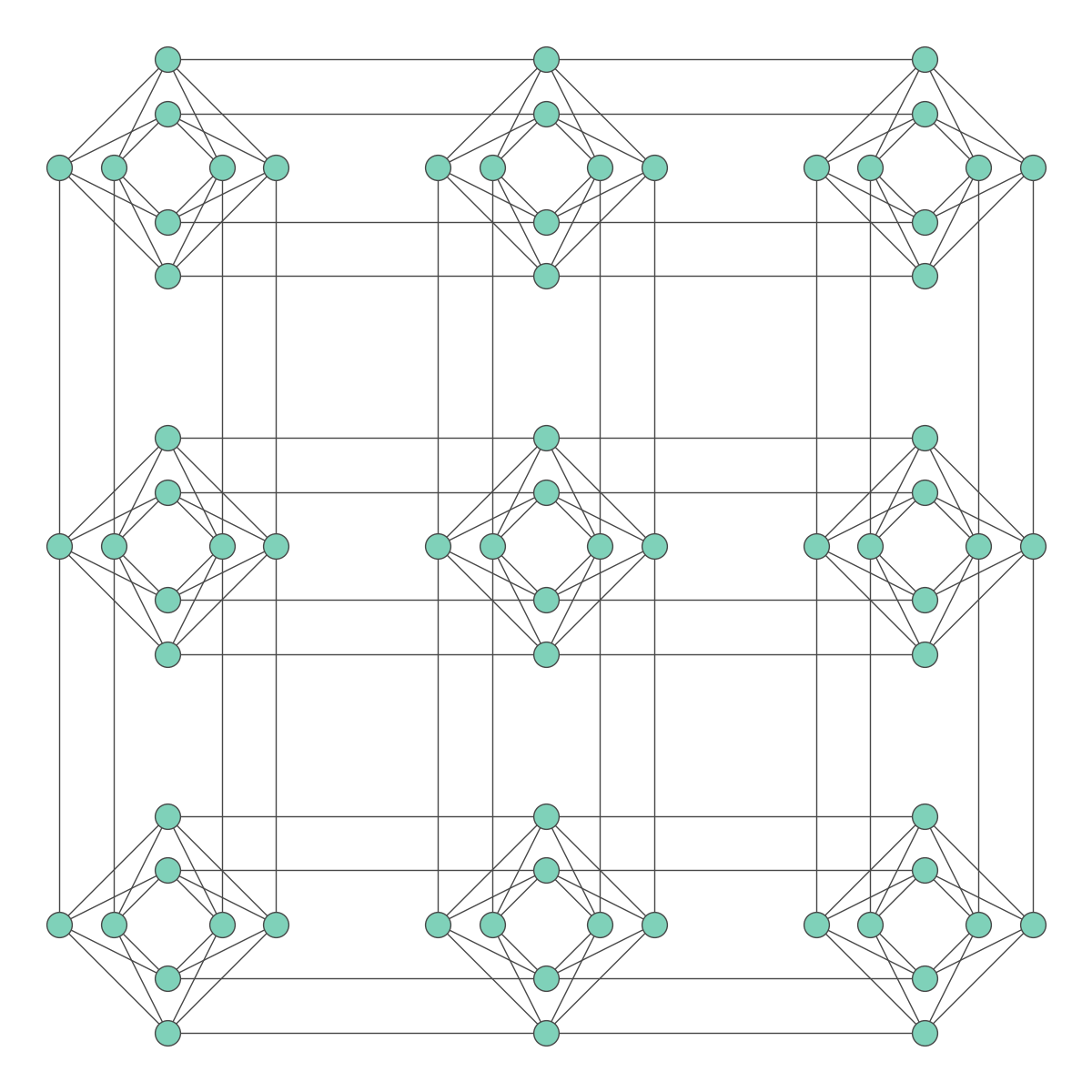}
\includegraphics[width=\linewidth]{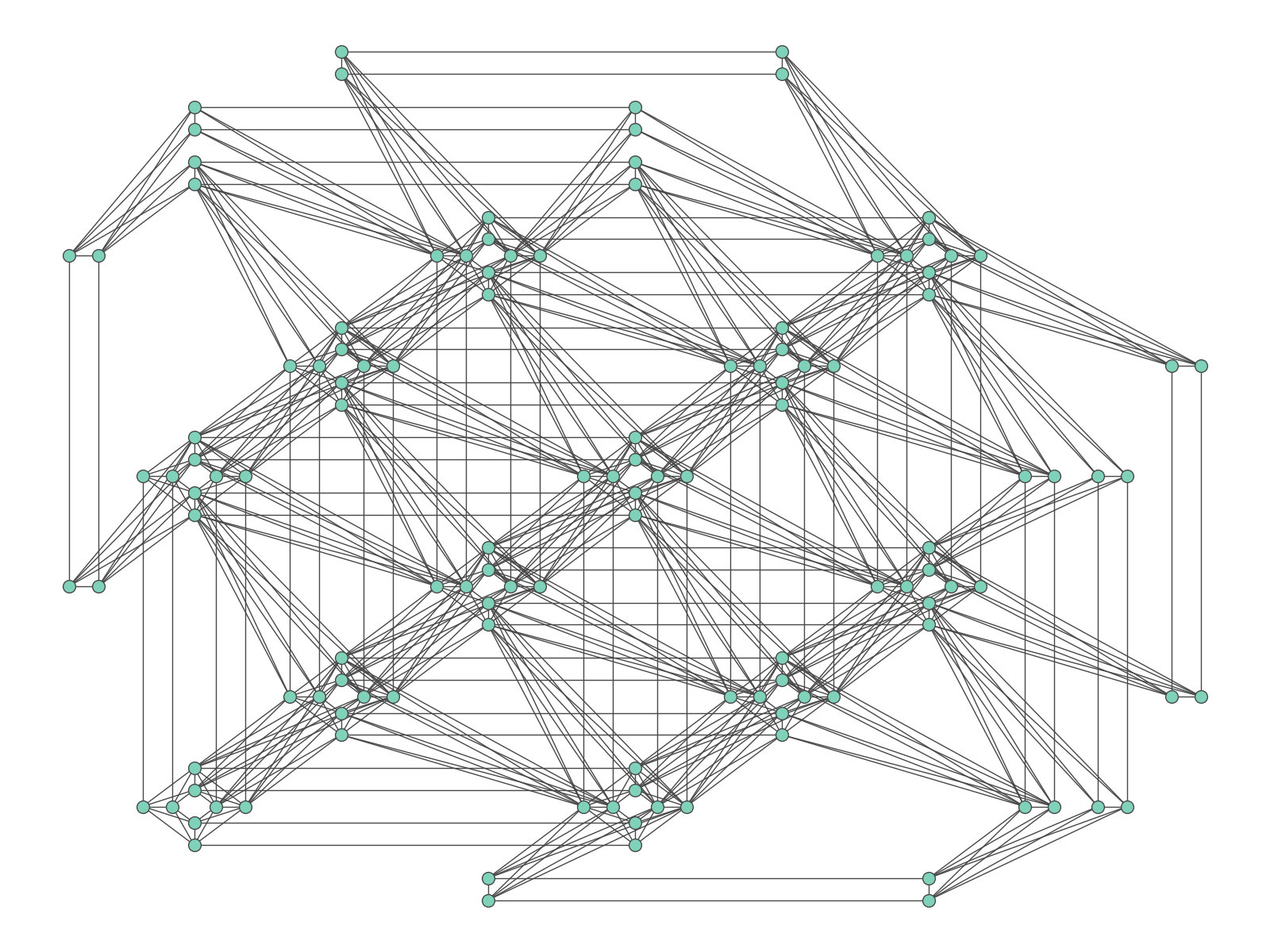}
  \caption{{\bf Top:} structure of the Chimera graph featured by DW2K. Shown is a so-called $C_3$ graph, i.e. a square with three cells per side. The pattern repeats in the same fashion to scale to larger chips. In particular, the device we use has a $C_{16}$.
  {\bf Bottom:} structure of the Pegasus graph of DWA. Shown is a $P_3$ graph. As with Chimera, the pattern repeats and the actual device has a $P_{16}$.
  Chimera is a subgraph of Pegasus.}
  \label{fig:architectures}
\end{figure}

\section{Problem mapping: minimum spanning tree}
\label{appendix:mapping_mst}

Consider a graph $G=(V,E)$ with weights $w(E)$ for each edge, from which we wish to obtain a minimal weighted spanning tree with maximum degree $\Delta$, i.e., find its BD MST. This involves minimizing the sum of the weights of the tree edges, represented by the cost function

\begin{align}
C_0 = \sum_{p,v} w_{pv}x_{p,v},
\label{eq:cost}
\end{align}
which we explain below. Several constraints are also imposed to ensure that the graph is in fact a spanning tree and its degree is bounded by $\Delta$.

A root for the tree is picked randomly or based on problem structure---generally, picking a high-degree vertex as the root will result in lower resource costs---and assigned to level $1$. Its children will be at level $2$, their children at level $3$ and so on, leading to the `level-based' designation.

The variables $x_{p,v}$ appearing in Eq.~\eqref{eq:cost} represent the parent-child relationships in the tree; $x_{p,v}=1$ if $p$ is the (adjacent) parent of $v$ (and $0$ if not). The indices $p,v$ range over $p=1,\dots,n$ and $v=2,\dots,n$, restricted to (intersected with) pairs $(p,v)$ or $(v,p)$ that occur in $E$. Thus there are two variables for every edge not containing the root, and one for every root edge, giving $2m-d_r$ total $x_{p,v}$ variables, with $m$ being the number of edges in $E$ and $d_r$ the degree of the root.

Since our problem needs to be in QUBO form, the constraints will be expressed as penalty terms. The first penalty term enforces that every node (except the root) has exactly one parent,
\begin{align}
 C_{pen1} = \sum_{v\in\{2,\dots,n\}}
\left(\sum_{p:(pv)\in E} x_{p,v}-1\right)^2\;.
\end{align}
The number of terms in the sum is $2m-d_r$, i.e. equal to the number of variables $x_{p,v}$.

The second penalty term enforces that each vertex exists at exactly one level in the tree,
\begin{align} 
C_{pen2} = \sum_{v\in\{2,\dots,n\}} 
\left(\sum_{\ell=2}^n y_{v,\ell}-1\right)^2\;.
\end{align}

It introduces the $y_{v,\ell}$ variables, with $y_{v,\ell}=1$ if $v$ is at depth $\ell$ of the tree, $v=2,\dots,n$, $\ell=2,\dots,n$. There are $(n-1)^2$ such variables. However, since the number of variables will eventually determine how many logical qubits the problem requires, it is in our interest to reduce it as much as possible. By picking the root smartly the range of $\ell$ can be reduced. We also carry out the following preprocessing: taking the original graph $G=(V,E)$, the distance from each node to the one we select as the tree root is calculated. Given that it is impossible for a node to be at a level smaller than its distance to the root, we can avoid generating any $y_{v,\ell}$ for which that is the case, further bringing down the total number of $y_{v,\ell}$ variables.

The third penalty term enforces that the tree has degree at most $\Delta$,
\begin{align}
C_{pen3} = &\sum_{p=2}^v \left(\sum_{v:(pv)\in E} x_{p,v}-\sum_{j=1}^{\Delta-1}z_{p,j}\right)^2 \nonumber \\
+ & \left( \sum_{v:(1v)\in E} x_{1,v} - \sum_{j=1}^{\Delta}z_{1,j} \right)^2.   
\end{align}
It is separated into two terms to account for the fact that the root can have up to $\Delta$ children, while all other nodes cannot have more than $(\Delta - 1)$, since they have a parent. 
To enforce the inequality $\sum_{v:(pv)\in E} x_{p,v} \le \Delta-1$, integer variable $z_p\in [0,\Delta-1]$ is introduced as slack variable, and the inequality is enforced as equality $\sum_{v:(pv)\in E}  x_{p,v} = z_p$.  The integer variable is further encoded into binary variables $z_{p,j}$.  In general, various encoding methods can be applied to encode an integer into binaries, including binary, unary, and one-hot encodings.  While binary encoding is most efficient for integers of value power of two, we use unary encoding here, which can be applied straightforward to an arbitrary value of $\Delta$.

The fourth and final penalty term enforces that the tree encoding is consistent, i.e., that if $p$ is the parent of $v$ then its level is one less than $v$'s, 
\begin{align}
 C_{pen4} = &\sum_{p,v} 
\sum_{\ell=3}^n x_{p,v} y_{v,\ell}(1-y_{p,\ell-1}) \nonumber \\ 
+ &\sum_{v=2}^{d_r} x_{1,v}(1-y_{v,2})
+ \sum_{v=2}^{d_r} y_{v,2}(1-x_{1,v})\;,
\end{align}
where the last two sums handle the edges connected to the root and their terms are quadratic, while the first sum deals with the remaining edges and produces cubic terms of the form $x_{p,v} y_{v,\ell}(1-y_{p,\ell-1})$. While the original number of cubic terms would be $$(2m-2d_r)*(n-2),$$ 
thanks to the preprocessing of the $y_{v,\ell}$ variables this number is reduced. 
Because cubic terms cannot be directly encoded in D-Wave, we introduce an ancilla variable
$a_{p,v,\ell}$ to encode  $x_{p,v} y_{v,\ell}$,
and accordingly a penalty function $f(x,y,a)=3a+xy-2ax-2ay$ is added to raise a penalty if $a=xy$ is violated.
The term $x_{p,v} y_{v,\ell}(1-y_{p,\ell-1})$ then can be replaced by quadratic terms

\begin{align} 4a - ay_{p,\ell-1}
+ x_{p,v} y_{v,\ell} - 2 ax_{p,v}- 2a y_{v,\ell}\;.
\end{align} 

The total number of variables (and hence, logical qubits) without preprocessing is at most:
\begin{align*}
  2m-d_r  + (n-1)^2 + n(\Delta-1)+1 + (2m-2d_r)(n-2) \\
  \simeq 2mn + n^2  
\end{align*}

This would mean, for instance, that the complete graph $K_5$ with $\Delta =3$ would require between $86$ and $100$ logical qubits (depending on $d_r$). With preprocessing, we are able to bring this number down to $74$.

Finally, we can write the overall objective function as
\begin{align} 
\label{eq:C_tot}
 C = C_0 + A(C_{pen1}+C_{pen2}+C_{pen3}+C_{pen4})\;,
 \end{align} 
and accordingly the cost Hamiltonian $H_C$.
In Eq.~\eqref{eq:C_tot} we define the minimum penalty weight to be the maximum edge weight
\begin{align} 
A= w_{max} + \varepsilon =\max_{(uv)\in E} w_{uv} + \varepsilon\;.
\end{align}

\section{Problem mapping: graph coloring}
\label{appendix:mapping_gc}

We use a standard one-hot encoding for the GC problem. We have an undirected graph $G = (V, E)$ with $n$ nodes, and a number of colors $k$. We define the binary variables:

\begin{equation}
  x_{v,i} =
    \begin{cases}
      1 & \text{if node $v$ is colored with color $i$}\\
      0 & \text{otherwise},
    \end{cases}  
\end{equation}

for a total of $n$x$k$ variables. Then our Hamiltonian is simply:

\begin{equation}
    H = \sum_{v=1}^n \left(1 - \sum_{i=1}^k x_{v,i} \right)^2 + \sum_{(uv) \in E} \sum_{i=1}^k x_{u,i} x_{v,i}.
\end{equation}

The first term enforces that each node only has one color assigned, while the second one penalizes any pair of nodes connected by an edge which have the same color. $H = 0$ if there is a $k$ coloring for graph $G$, and $H > 0$ otherwise.

\section{Problem mapping: information sharing}
\label{appendix:mapping_info}

\subsection{Problem inputs}
\label{appendix:info_problem_inputs}

Given the following:
\begin{itemize}
\item A discrete time variable $t$, taking the non-negative integer values.
\item A finite set $I = \{0, 1, \ldots, n-1\}$, whose elements each represent a message to be transmitted over a communications network, modeled below.
\item A graph $G = (V, E)$, which models a communications network.  All messages $i \in I$ are sent from the same node, called the {\em sender node} and denoted $v_{{\footnotesize sender}} \in V$.  Each sent message $i$ is intended to reach, eventually, its {\em recipient node} $r_{i} \in V$.  (Different messages generally have different recipients.)
\item  A {\em transmission time function} $l$ that assigns to each pair $(i, e)$, where $i \in I$ is a message and $e \in E$ an edge in the communications graph $G$, a time duration (a non-negative integer) $l(i, e)$, which models how long message $i$ will take to traverse edge $e$.  The latter value, when it is necessary to indicate the constituent vertices $(v, v')$ of $e$, can be written in the more detailed notation $l(i, v, v')$.

\item A {\em transmission of a message along a path $P$}, which is a triple

\begin{equation}
\label{equation:transmission-of-message}
T(i) = \big(i, t^{em}_{i}, P = (v_{0} = v_{{\footnotesize sender}}, v_{1}, \ldots, v_{N_{i}} = r_{i})\big),
\end{equation}

where $i \in I$ is a message, $t^{em}_{i}$ is the {\em emission time} of the message (i.e., the instant when the message is sent), and $P$ is a path in $G$ from the sender node to the recipient node of message $i$.  

The set of all possible transmissions of message $i$, along all possible paths $P$ and with all possible emission times, will be denoted by ${\cal T}^{i}$.
It follows that once a message transmission (\ref{equation:transmission-of-message}) has succeeded\footnote{It can fail because of the edge capacity limitations, defined below.}, its total duration is given by
\begin{equation}
D(i,P) := \sum_{k = 0}^{ N_{i}-1} l(i, v_{k}, v_{k+1}),
\end{equation}
\label{eq:message_travel_time}
and that message $i$ arrives at its recipient node at time
\begin{equation}
t^{em}_{i} + D(i,P).
\end{equation}

The partial sums from $D(i,P)$ correspond to the times at which message $i$ enters a given edge in path $P$.  In more detail, under transmission (\ref{equation:transmission-of-message}), edge $(v_{j}, v_{j+1})$ is entered by message $i$ at time
\begin{equation}
t^{em}_{i} + \sum_{k = 0}^{j} l(i, v_{k}, v_{k+1}).
\end{equation}

\item A scheduled time of emission for each message $i \in I$,
\begin{equation}
t^{sched}_{i}.
\end{equation}

In no transmission  (\ref{equation:transmission-of-message}) of that message can the actual $t^{em}_{i}$ emission time occur before $t^{sched}_{i}$; thus, one of the constrains is the inequality

\begin{equation}
\label{equation:emission-time-constraint}
t^{em}_{i} \geq t^{sched}_{i}.
\end{equation}

The non-negative difference 

\begin{equation}
t^{em}_{i} -  t^{sched}_{i}
\end{equation}

is called {\em the delay of transmission  (\ref{equation:transmission-of-message})}.

\item  A {\em transmission schedule}; a mapping $S$ that assigns to each message $i \in I$ a transmission:
\begin{equation}
S : i \mapsto T(i) \in {\cal T}^{i}.
\end{equation}

The set of all possible transmission schedules is the Cartesian product
\begin{equation}
{\cal S} = {\cal T}^{0} \times {\cal T}^{1} \times \ldots \times {\cal T}^{n-1}.
\end{equation}

\item A {\em cost of transmission delay} for each message $i \in I$, $c_i$, that represents how much it costs to delay a message per unit of time, and can be interpreted as a general way of expressing how important it is for the message to arrive promptly.

\item  A {\em node capacity function $B$}. Each node in $G$ is assumed capable of allowing entry to no more than some finite number of messages.  This number will be called the {\em capacity} of the node. In some cases, a message can have such high priority that its arrival without delay must be guaranteed (i.e. a message $i$ such that $c_i \gg c_j \forall j \neq i$). The path of such a message is initially hardcoded into the problem through the node capacity function $B$, with the addition of time dependence (the capacity is reduced at the nodes and times corresponding to the message's path). Thus, the node capacity function $B$ (treated as a discrete control variable) assigns to each node $v$ and at each time instant $t \in \{0, 1, \ldots \}$ a non-negative integer
\begin{equation}
B(t, v) \equiv B^v_t.
\end{equation}

\end{itemize}

The {\bf Information Sharing Problem} is to find a transmission schedule 
\begin{equation}
S = (T(0), T(1), \ldots, T(n-1)) \in {\cal S}
\end{equation}
that minimizes the total cost of delay
\begin{equation}
\sum_{i \in I} c_i \left(t^{em}_{i} -  t^{sched}_{i}\right), 
\end{equation}
subject to the constraints (\ref{equation:emission-time-constraint}) and
\begin{equation}
\left(\begin{array}{c}
\mbox{number of messages}\\
\mbox{entering node $v$}\\
\mbox{at time $t$}\\
\mbox{under schedule $S$}\\
\end{array}
\right)
\leq B^v_t \quad 
\left\{
\begin{array}{c}
\mbox{for all node entry}\\
\mbox{times $t$ involved in}\\
\mbox{one or more of the}.\\
\mbox{transmissions $S(i)$}.\\
\end{array}
\right. 
\end{equation}

\subsection{Mapping to QUBO}
\label{appendix:info_qubo}

We use a quantum annealer to solve instances of the information sharing problem where, given a path $P_i$ for each message $i \in I$, an emission time $t^{em}_i$ is assigned to each message such that the total cost of delays is minimized. For simplicity, the scheduled time of emission $t^{sched}_{i}$ is taken to be 0 $\forall i \in I$. 

The problem must be mapped to QUBO to be solved on the quantum annealer. The objective function to minimize will be:

\begin{equation}
    C_{obj} = \sum_{i \in I} c_i a_i,
\end{equation}

where $a_i$ is the total delay for message $i$, i.e. the actual arrival time minus the scheduled arrival time:

\begin{equation}
    a_i = \sum_{t_a} t_a x_{i,{t_a}}^{r_i} - D(i, P_i),
\end{equation}

where the sum is over the possible arrival times $t_a$, and $D(i, P_i)$ is given by Eq.~\ref{eq:message_travel_time}.

In the expression for the total delay, we use the binary variables:

\begin{equation}
  x_{i,t}^v =
    \begin{cases}
      1 & \text{if message $i$ arrives at node $v$ at time $t$}\\
      0 & \text{otherwise},
    \end{cases}  
\end{equation}

which means that for given $i$ and $v$, $x_{i,t}^v=1$ for a single value of $t$ and 0 for all others.

We have three constraints that are expressed as penalties to formulate a QUBO. The first penalty term enforces that the capacity of the network is not exceeded:

\begin{equation}
    C_{pen1} = \sum_{t=1}^{t_h} \sum_{v \in P_i} \left( \sum_{i \in I} x_{i,t}^v - \sum_{k=1}^{B_t^v} s_{k,t}^j \right)^2, 
\end{equation}

where the $s_{k,t}^v \in [0, 1]$ are slack variables, and $t_h$ is the {\it time horizon}, a limit on the latest time we consider for all the messages to have been transmitted, which is needed for practical purposes (the sum needs an upper limit).
If $\sum_{i \in I} x_{i,t}^v \leq B_t^v$ is satisfied, there exists a variable assignment for the $s_{k,t}^v$ that makes $C_{pen1} = 0$.

The second penalty term enforces path connectivity is respected, by ensuring that a message does not arrive at a node faster than the travel time between that node and the previous one in its path would allow. This penalty term counts the number of path connectivity violations:

\begin{equation}
    C_{pen2} = \sum_{i \in I} \sum_{v \in P_i} \sum_{t=1}^{t_h-l_{v,v+1}^i} \sum_{k=0}^{l_{v,v+1}^i -1} x_{i,t+k}^{v+1} x_{i,t}^{v}.
\end{equation}

The third and last penalty term makes sure that each message arrives at each of the nodes in its path exactly once; this ensures that the message is actually transmitted and travels through its path as it should:

\begin{equation}
    C_{pen3} = \sum_{i \in I} \sum_{v \in P_i} \left(\left(\sum_{t=1}^{t_h} x_{i,t}^{v}\right) - 1\right)^2.
\end{equation}

If we generated every variable regardless of feasibility (i.e., whether there exists a valid scenario in which the variable is nonzero), we would have $n$x$|E|$x$t_h$ $x_{i,t}^v$ variables. We can reduce this number with two considerations about feasibility. First, we need only to generate variables for the nodes in each message's path, but not the rest. This reduces the $n$x$|E|$ to $\prod_{i=0}^{n-1} N_i$. Second, we can avoid times that are too early, by precalculating the earliest time that each message $i$ can arrive at node $v_j \in P_i$:

\begin{equation}
    t_{e,v_j}^i = \sum_{j'=0}^{j-1} l(i, v_{j'}, v_{j'+1}).
\end{equation}

Then, instead of generating variables for the time interval $[1, t_h]$ we can do it for $[t_{e,v_j}^i, t_h]$ for each $i \in I$ and $v_j \in P_i$. With this, the total number of $x_{i,t}^v$ variables is $\sum_{i \in I} N_i t_{e,v}^i$. 

With this, the sum over time in the second penalty term $C_{pen2}$ would start at $t = t^i_{e,v^i_1}$ rather than $t = 1$, where $v_1$ is the second node in the path of message $i$. We can start this sum at the earliest time that the message can be in its second path node, because the variables corresponding to the first path node are created to respect the minimum travel time, and thus there is no need to double check them. For the third penalty term $C_{pen3}$, the sum over time can start at $t = t^i_{e,v \in P_i}$ instead of $t=1$.

Finally, the number of slack variables is $\prod_{v \in \bigcup\limits_{i\in I} P_i} \prod_{t=1}^{t_h} B_t^v$, and all need to be generated. The first penalty term $C_{pen1}$ is unaffected by the binary variable pruning described in the previous paragraph; all times beginning at $t=1$ need to be checked for bandwidth violations, given that these are unrelated to minimum travel times.

Taking all of the above into consideration, the full Hamiltonian for our QUBO problem is:

\begin{equation}
    H = C_{obj} + A_1 C_{pen1} + A_2 C_{pen2} + A_3 C_{pen3},
\end{equation}

where $A_i$ are the penalty weights. Penalty weights are needed because without them it could be advantageous to violate a penalty term, by virtue of such a penalty being smaller than the cost reduction it affords. To avoid this scenario, we need to consider the maximum cost reduction that could be obtained by each penalty violation, and then choose a penalty weight that will ensure the penalty is always larger than said cost reduction.

Let us start with the first penalty term $C_{pen1}$, which enforces bandwidth. At each time and each node in a path, surpassing bandwidth by one message results in a $+1$ penalty before multiplying by $A_1$. Additional messages would cause an even larger penalty due to the square, so it is sufficient to consider the first one. The largest possible reduction to $C_{obj}$ without violating any additional penalties would be the maximum, over all messages, of a message going from arriving at the latest possible time to the earliest possible: $\text{max}_i (c_i (t_h - D(i, P_i))$. A further reduction could be attained by arrival at an earlier time than physically possible (i.e. smaller than $D(i, P_i)$), but this would result in an additional penalty by either not arriving at all (reflected in $C_{pen3}$) or not respecting path connectivity ($C_{pen2}$). It is then sufficient to choose $A_1 = \text{max}_i (c_i (t_h - D(i, P_i))) + \varepsilon$.

The second penalty term $C_{pen2}$ enforces path connectivity, through pairs of $x$ variables that cannot be 1 at the same time because they correspond to subsequent path nodes of a given message at times closer than the minimum travel time between said nodes. Each such pair where both variables are 1 results in a $+1$ penalty before multiplying by $A_2$. The maximum $C_{obj}$ reduction that can be obtained from a single pair occurs when traveling is instantaneous, i.e. the message is early by some time $l(i, v_j,v_{j+1})$. So the penalty weight can be set to $A_2 = \text{max}_{i,j} c_i l(i, v_j,v_{j+1}) + \varepsilon$. Note that this is never larger than the overall maximum reduction to $C_{obj}$ described in the previous paragraph ($\text{max}_i (c_i(t_h - D(i, P_i)))$), given that $t_h \geq D(i, P_i)$ and $D(i, P_i) \geq l(i, v_j,v_{j+1})$ (see Eq.~\ref{eq:message_travel_time}).

\begin{figure}[h]
\includegraphics[width=\linewidth]{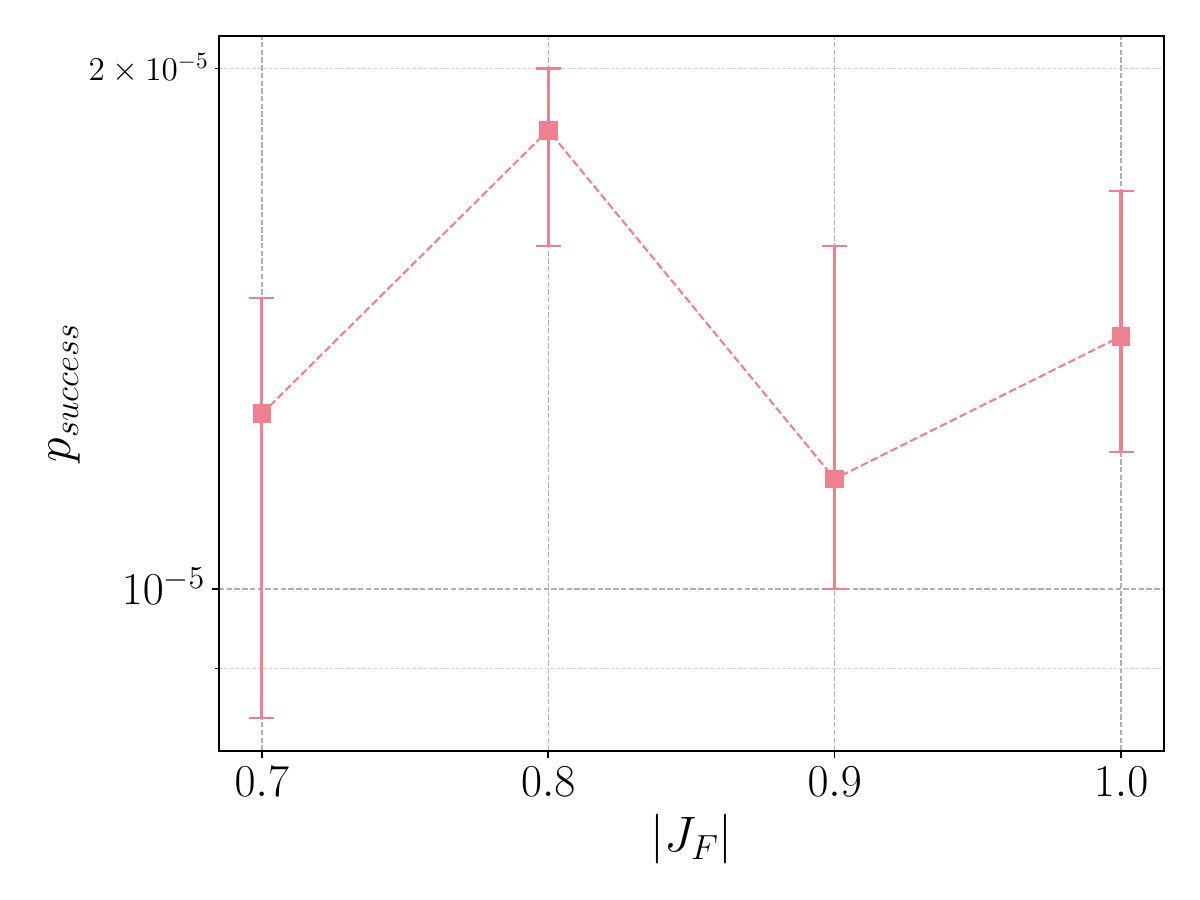}
  \caption{{\bf Optimal $|J_F|$ for ensemble of 50 $n=6$ BD MST instances (DWA).} 
  }
  \label{fig:p_vs_jf_mst_n6}
\end{figure}

\begin{figure}[h]
\includegraphics[width=\linewidth]{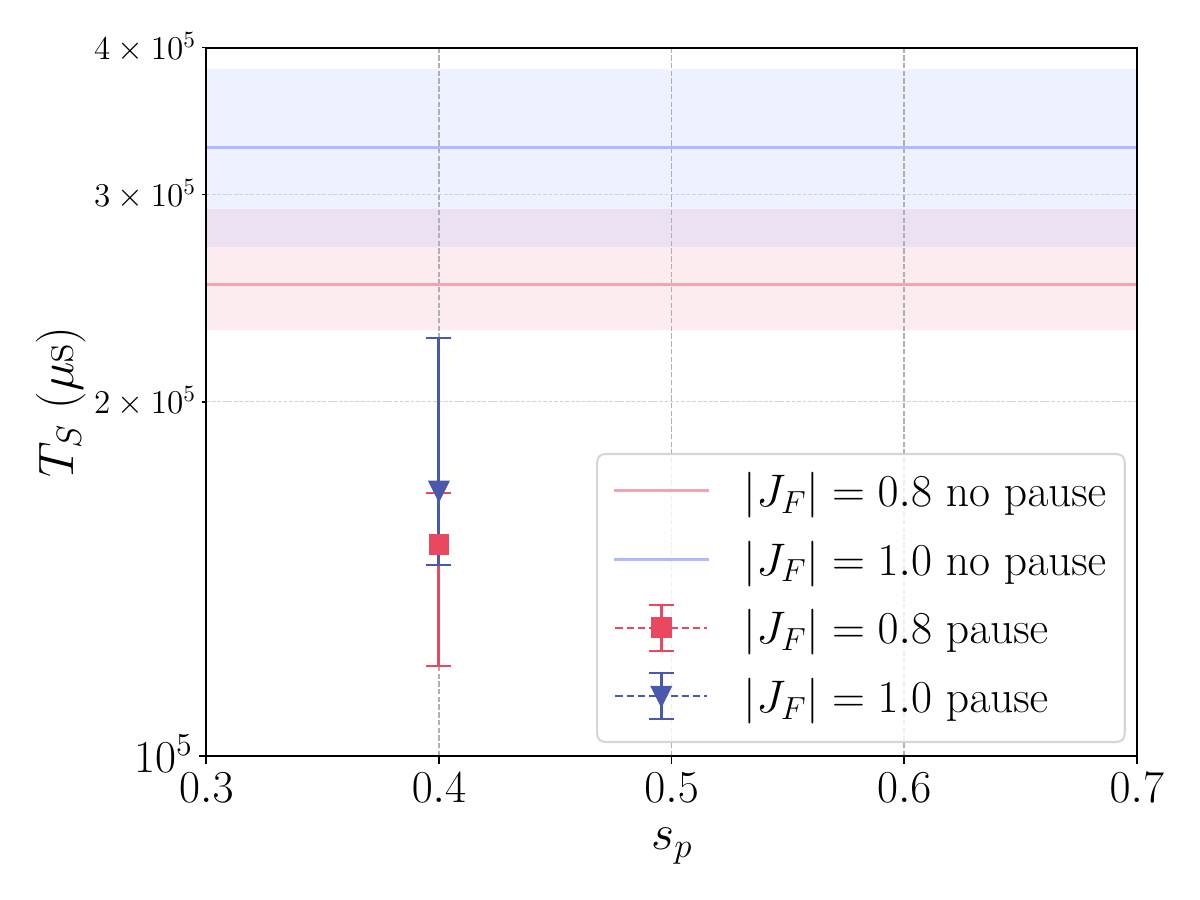}
  \caption{{\bf Improvement of $T_S$ with pause for ensemble of 50 $n=6$ BD MST instances (DWA).} 
  }
  \label{fig:tts_vs_sp_mst_n6}
\end{figure}

Finally, the third penalty term $C_{pen3}$ involves arrivals. A penalty of +1 (before multiplying by $A_3$) is incurred every time a message does not arrive to one of its path nodes. A $C_{pen3}$ violation has the potential for the greatest reduction in $C_{obj}$, in the case where a message should arrive at its destination at time $t_h$, and it simply does not arrive, reducing $C_{obj}$ by $c_i t_h$. Then, we need to set $A_3 = \text{max}_i c_i t_h + \varepsilon$.

Because this last penalty weight is the highest of the three, if we want to set a single penalty weight for all penalties that will suffice in any situation, $A_1 = A_2 = A_3 = \text{max}_i c_i t_h + \varepsilon$ is the appropriate choice.

For our particular set of instances, we empirically find a lower penalty weight of $A_1 = A_2 = A_3 = 1/2 \sum_{i \in I} c_i \sum_{t=t_{a_{min}}}^{t_h} t + \varepsilon$ to be sufficient, while helping keep coefficients at a similar scale to avoid running into precision issues with the hardware.

\section{Optimizing $t_a$ for different problem classes}
\label{appendix:ta_opt}

\begin{figure*}[ht]
\includegraphics[width=0.32\linewidth]{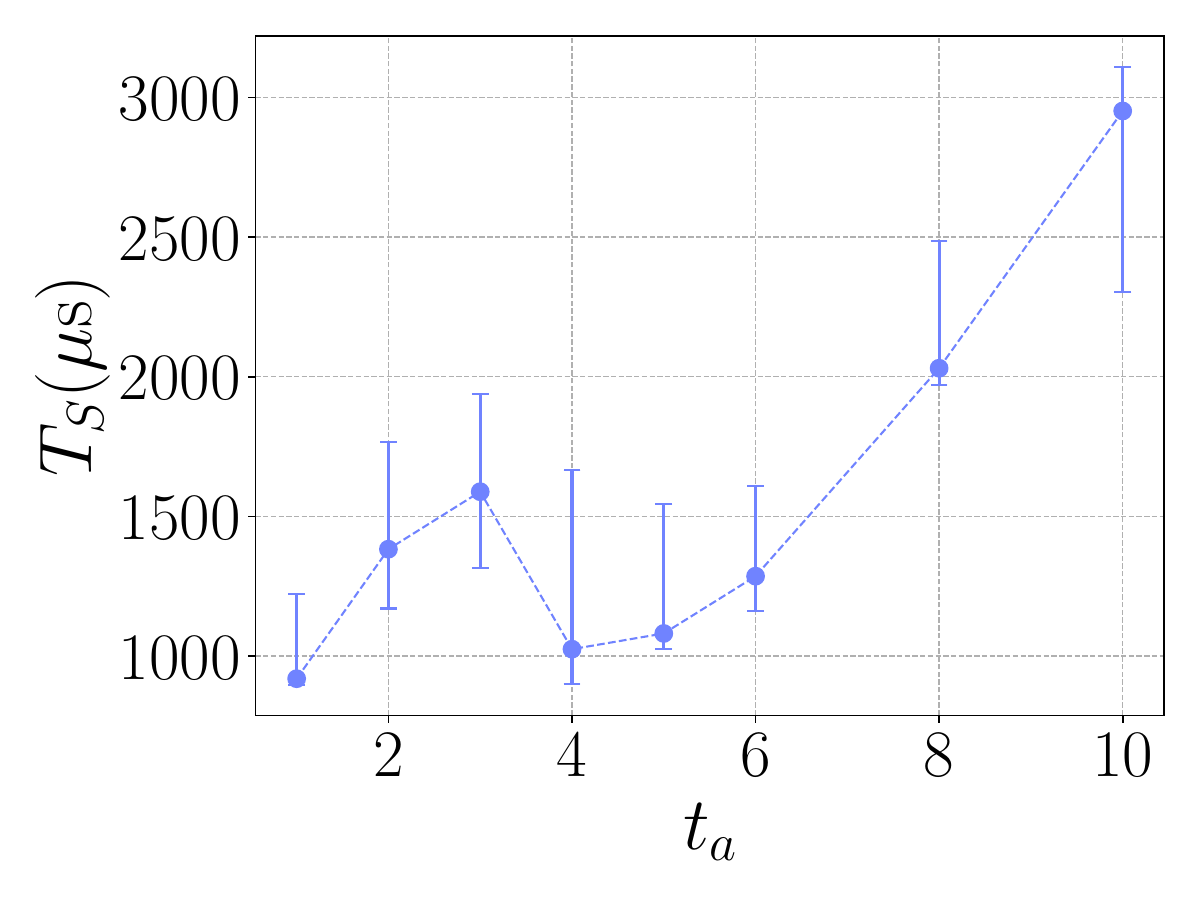}
\includegraphics[width=0.32\linewidth]{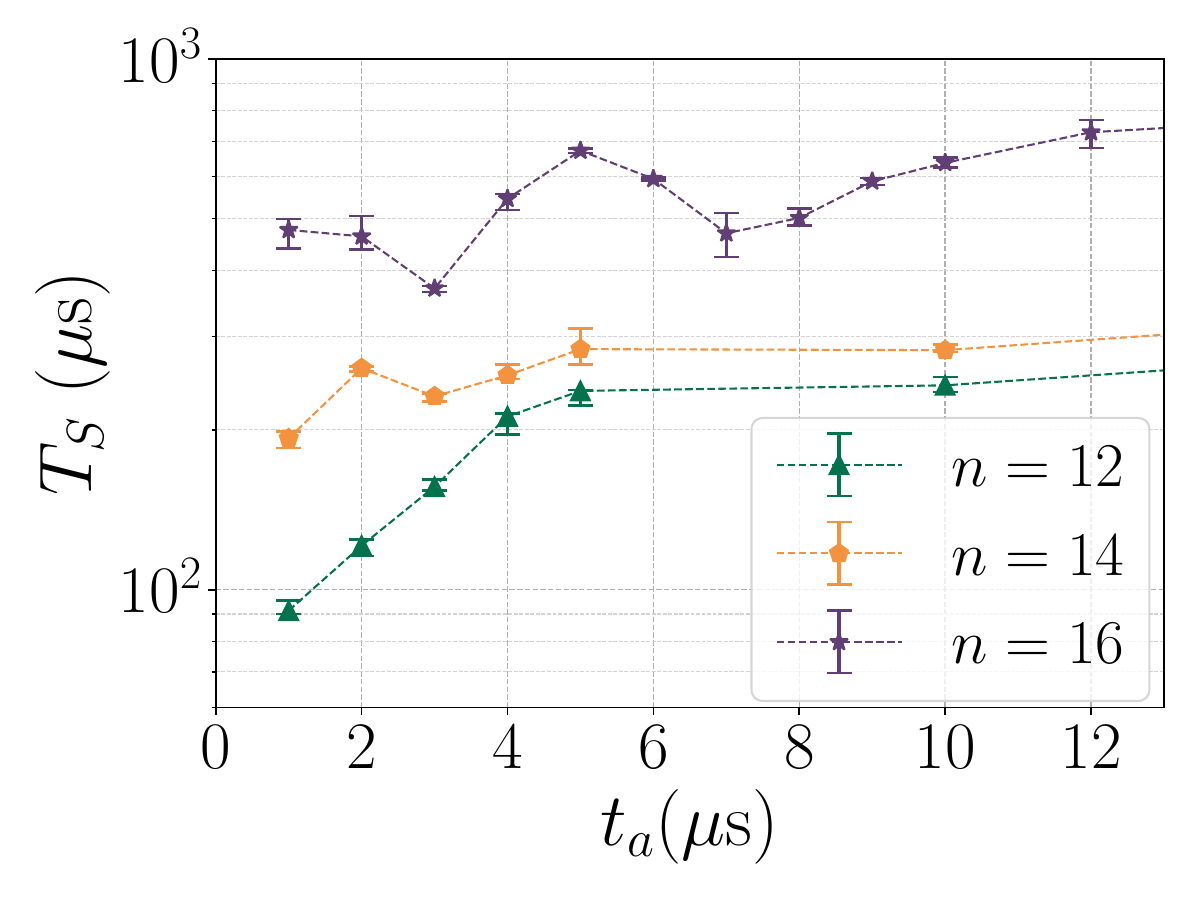}
\includegraphics[width=0.32\linewidth]{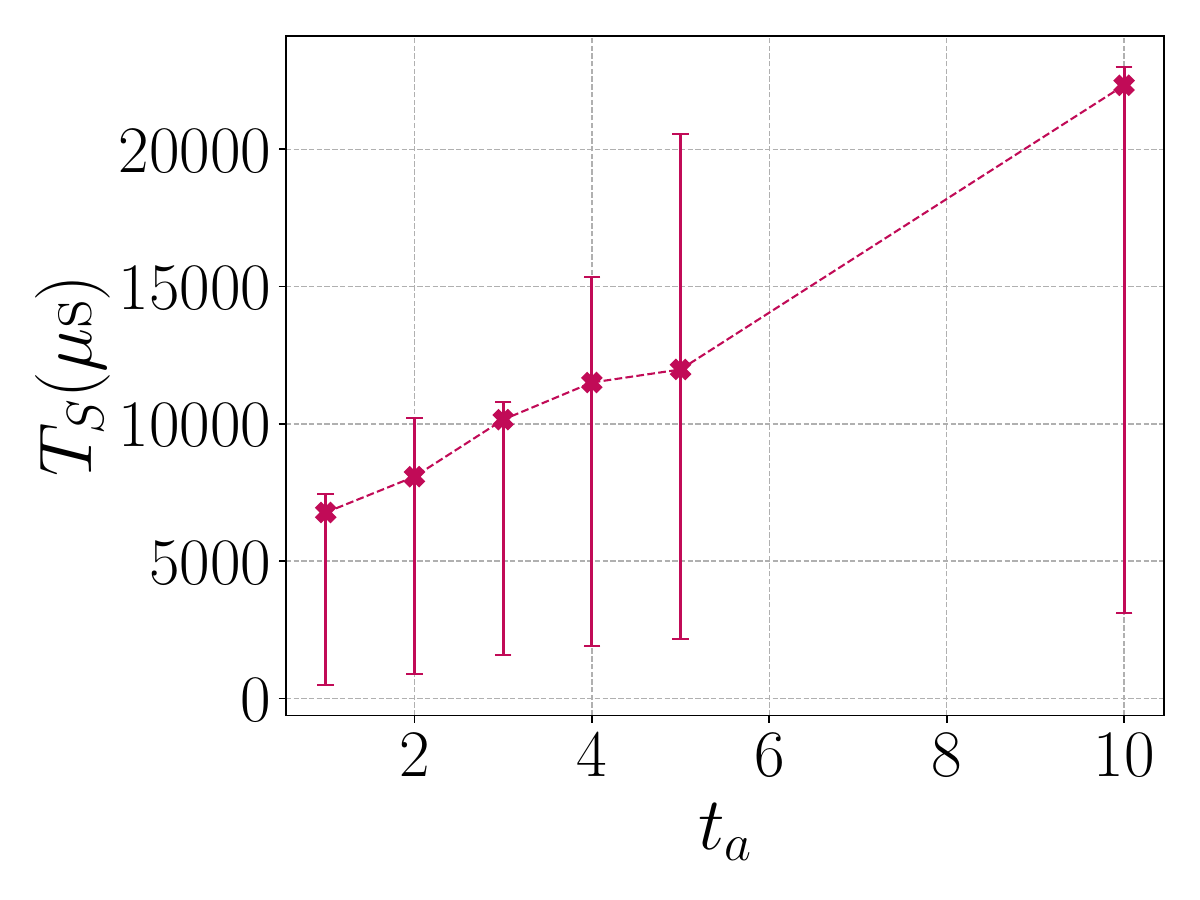}
  \caption{{\bf $T_S$ versus $t_a$ for the three different problems run on DWA.} Data points show median bootstrapped over instances, with error bars at 35th and 65th percentiles. {\bf Left:} BD MST, 45 $n=5$ instances. {\bf Middle:} GC, 3 ensembles of 20 instances each, with legend showing their respective $n$. {\bf Right:} INFO, nine instances.}
  \label{fig:ta_optimal_all_problems}
\end{figure*}

As seen in Fig.~\ref{fig:ta_optimal_all_problems}, the optimal $t_a$ is $\leq 1 \mu$s in most cases. 

For the BD MST ensemble there is a local minimum at $t_a = 4 \mu$s with $T_S$ almost as good as that obtained at 1 $\mu$s. This effect is consistent even if we look at the instances individually; the low $T_S$ at 4 $\mu$s is not a result of some instances preferring a shorter $t_a$ and others a longer one, but instead of many of them having a curve similar to that of the ensemble.

For GC, the three ensembles separated by size tell a more complete story. The best $T_S$ in the accessible range for $n=12$ is clearly at $t_a = 1 \mu$s, with longer times doing progressively worse. This holds true for $n=14$ as well, but a local minimum at 3 $\mu$s appears. And finally, at $n=16$, the minimum at 3 $\mu$s is the global one, with another one at 7 $\mu$s a close second. Both yield clearly better results than 1 $\mu$s, although it is possible that the true global minimum could be $< 1 \mu$s. Because these results were somewhat noisy, we increase the number of anneals and instances to get more reliable data. For all $n$, the data points for $t_a = 1$, 2 and 3 $\mu$s are obtained with 500 gauges instead of our standard 100, for a total of 250,000 anneals. In the case of $n=16$, 500 gauges are also used for the $t_a = 4 \mu$s point, and all the data points use 30 instances instead of the usual 20. 

The optimal $t_a$ is also $\leq 1 \mu$s for the INFO ensemble. While error bars are very large, that is a product of there being only nine instances of disparate hardness, rather than lack of consistency in the results. Eight of the nine instances have 1 $\mu$s as their optimal within the device's range, with the $9^{\text{th}}$ one doing better at 2 $\mu$s. For six out of the nine, $T_S$ monotonically increases with $t_a$.

\section{Results for $n=6$ BD MST instances}
\label{appendix:mst_n6}

We also extend our results to an ensemble of 50 $n=6$ instances. These are much harder than the $n=5$ and, when a few were attempted on DW2K, none of them could be solved. Instead of 50,000 reads, we use $10^6$ reads to obtain good statistics. As shown in Fig.~\ref{fig:p_vs_jf_mst_n6}, $|J_F|=0.8$ still appears to yield the best $T_S$, although it is not as clear as for the $n=5$ ensemble. Due to the costly nature of solving these instances, we do not exhaustively explore different schedules as we did for the previous problems, but are able to verify that a pause of $t_p=0.2 \mu$s at $s_p=0.4$ improves upon the no pause results, which can be seen in Fig.~\ref{fig:tts_vs_sp_mst_n6}. Without a pause, 11 of the 50 instances did not solve after $10^6$ reads. With the pause, 10 out of 50 did not solve. It is possible that some of them would solve at a different $|J_F|$, given that a limited range is tested and, as we see for the $n=5$ ensemble, particularly difficult instances tend to require a stronger $|J_F|$.

We are able to confirm that for these difficult instances, our embedding method of picking the smallest (in number of physical qubits) out of ten leads to a significant improvement in the results. For $|J_F|=1.0$, if we simply use the first embedding we generate, we find $p_{success} =$ \SI{7e-6}, with $35^{th}$ and $65^{th}$ percentile values at \SI{6e-6} and \SI{8e-6} respectively. By generating ten embeddings and picking the smallest, we increase to $p_{success}=$ \SI{1.4e-5}, with $35^{th}$ and $65^{th}$ percentiles at \SI{1.2e-5} and \SI{1.7e-5}, a $2\times$ improvement.

\end{document}